\documentclass[a4paper]{spie}  
\usepackage[]{graphicx}
\usepackage{apj-jour}

\title{Laue gamma-ray lenses for space astrophysics: status and prospects} 

\author{Filippo Frontera\supit{1} and Peter von Ballmoos\supit{2}, 
\skiplinehalf
\supit{1}University of Ferrara, Physics Department, Via Saragat 1, 44100
Ferrara, Italy; \\
\supit{2}Centre d'Etude Spatiale des Rayonnements, 9, Avenue du Colonel Roche,
31028 Toulouse, France\\
}

\authorinfo{Further author information: (Send correspondence 
to F.F.)\\ F.F: E-mail: frontera@fe.infn.it, Telephone: +39 0532 974 254}

\begin{document} 
  \maketitle 

\begin{abstract}
We review  feasibility studies, technological developments and
the astrophysical prospects for Laue lenses devoted to
hard X-/gamma-ray astronomy observations. 

\end{abstract}

\section{INTRODUCTION}
\label{s:intro}  

Hard X-/soft gamma-ray astronomy is a crucial  window for the study of the most energetic and 
violent events in the Universe.  With the ESA INTEGRAL observatory \cite{Winkler03}, 
and the NASA {\em Swift} satellite \cite{Gehrels04}, unprecedented sky surveys 
in the band beyond 20 keV are being performed \cite{Bird10,Cusumano09}. 
As a consequence, hundreds of celestial sources have already been discovered, 
new classes of Galactic sources are being identified, an overview of the 
extragalactic sky is available, while evidence of extended matter-antimatter annihilation
emission from our Galactic  center \cite{Weidenspointner08} and of Galactic
nucleosynthesis processes have been also reported \cite{Weidenspointner08,Diehl06}.
However, in order to take full advantage of the extraordinary potential of soft 
gamma--ray astronomy, a new generation of telescopes is needed.
The current instrumentation  has relied on the use of direct--viewing detectors with
mechanical collimators (e.g., BeppoSAX/PDS, Ref.~\citenum{Frontera97}) and, in some
cases, with modulating aperture systems, such as  coded masks (e.g., INTEGRAL/IBIS, 
Ref.~\citenum{Ubertini03}).
These telescopes are penalized by their modest sensitivities, that improve at best as 
the square root of the detector
surface.  The only solution to the limitations of the current generation 
of gamma--ray instruments is the use of focusing optics. To study either the
continuum  emission or the nuclear line emission from celestial sources,
Laue lenses, based on diffraction from crystals in a transmission configuration,
are particularly suited to focus photons in the hard X--/soft gamma--ray ($<$ 1 MeV)
domain. As we will show, they show imaging capabilities for on-axis sources.

With these lenses, we expect a big leap in both flux sensitivity and angular resolution.
As far as the sensitivity is concerned, the expected increase is by a factor of at least 
10--100 with respect to the best non-focusing instruments of the current generation, 
with or without coded masks.
Concerning the angular resolution, the increase is expected to be more than a factor 10 
(from $\sim 15$ arcmin of the mask telescopes like INTEGRAL IBIS to 
less than 1 arcmin).

The astrophysical issues that are expected to be solved with the advent
of these telescopes are many and of fundamental importance. A thorough discussion of the science 
case has been carried out  in the context of the mission proposal {\em Gamma Ray Imager} (GRI), 
submitted to ESA in response to the first AO of the 'Cosmic Vision 2015--2025' plan 
\cite{Knodlseder07} (but see also Refs.~\citenum{Frontera05a,Frontera06,Knodlseder06}). 
We summarize here some of these issues.

\begin{itemize} 
\item {\em Deep study of high energy emission physics in the presence of super-strong 
magnetic fields (magnetars)}

The XMM and INTEGRAL observed spectra of Soft Gamma Ray
Repeaters \cite{Gotz06} and Anomalous X--ray pulsars \cite{Kuiper06}
leave unsolved the question of the physical origin of the high energy
component ($>$100 keV). A better sensitivity at E$>$100 keV is needed.
\item {\em Deep study of high energy emission physics in compact Galactic objects and  AGNs}

A clue to
the emission region and mechanism, along with the properties of the hidden black hole,
can be obtained with the measurement of the high energy cutoff and its relation with
the power--law energy spectrum of the compact objects. The current observational status is
far from clear ( see, e.g., Ref.~\citenum{Perola02,Risaliti02}). Much more sensitive observations
are needed, for both AGNs and compact Galactic sources. In the case of blazars, the
gamma--ray observations are crucial for the determination of their emission properties
given that their energy emission peaks at hundreds of keV \cite{Ghisellini09}.
\item {\em Establishing the precise role of non-thermal mechanisms in extended objects 
like Galaxy Clusters}

The existence of hard tails from Galaxy Clusters (GC) is still matter of discussion 
\cite{Rephaeli08}, and, should they exist, their origin is also an open issue.
Are they the result of a diffuse emission or are they due to AGNs in the GC? In the former case, 
what is the emission mechanism? What is their contribution to CXB? To answer these questions will
require much more sensitive observations, like those achievable with broad band Laue lenses.
\item {\em Origin of Cosmic hard X/soft gamma-ray diffuse background} 

Currently, a combination of unobscured, Compton thin and Compton thick radio-quiet AGN 
populations with different photon index distributions and fixed high energy
spectral cutoff ($E_c$)  are 
assumed in synthesis models of the Cosmic X--ray background (CXB) \cite{Gilli07}. Is it reasonable to 
assume a fixed $E_c$ for these sources? 
A photon-energy dependent contribution from radio-loud AGN to CXB, like blazars,  
is generally assumed. But their real contribution is still matter
of discussion. Deep spectral measurements of a significant sample of  AGNs
beyond 100 keV is needed to solve these issues. 
\item {\em Positron astrophysics}

Positron production occurs in a variety of cosmic explosion and acceleration sites, and the observation 
of the characteristic 511 keV annihilation line provides a powerful tool to probe plasma 
composition, temperature, density and ionization degree. The positron annihilation signature is 
readily observed from the galactic bulge region yet the origin of the positrons remains mysterious. 
Compact objects - both galactic and extragalactic - are believed to release significant numbers 
of positrons leading to 511 keV gamma-ray line emission in the inevitable process of annihilation. 
A recent SPI/INTEGRAL all-sky map \cite{Weidenspointner08} of galactic e$^-$e$^+$ annihilation radiation 
shows an asymmetric distribution of 511 keV emission that has been interpreted as a signature of 
low mass X-ray binaries with strong emission at photon energies $>$20 keV (hard LMXBs). A claim for 
an annihilation line  from a compact source (Nova Muscae) was reported in the 90s \cite{Goldwurm92} 
but was never confirmed. Much more sensitive observations are needed to study the 
annihilation line origin, sources and their nature.
\item {\em Physics of supernova explosions}

Type Ia supernovae (SNe Ia) are major contributors to the production of heavy elements and hence a critical 
component for the understanding of life cycles of matter in the Universe and the chemical evolution of 
galaxies. Because Laue lens telescopes allow the direct observation of radioactive isotopes that power 
the observable light curves and spectra, gamma-ray observations of SNe Ia that can be performed with this 
type of instrument are in a position to allow a breakthrough on the detailed physical understanding of 
SNe Ia. This is important for its own sake, but it is also necessary to constrain systematic errors when using 
high-z SNe Ia to determine cosmological parameters.

High resolution gamma-ray spectroscopy provides a key route to answering these questions by studying 
the conditions in which the thermonuclear explosion starts and propagates. 
A sensitivity of $10^{-6}$~photons cm$^{-2}$~s$^{-1}$ to broadened gamma-ray lines allows observations of 
supernovae out to distances of 50--100 Mpc. Within this distance it is expected that there will always be a 
type Ia SN in the phase of gamma-ray line emission, starting shortly after explosion, and lasting several months.

\end{itemize}

In this paper, we review the physical principles of Laue lenses, their geometry,
their optimization criteria, their optical properties, the current development status 
and the prospects for future missions for gamma--ray astronomy. 

%
%
\begin{figure}[htbp]
\begin{center}
\includegraphics[width=0.8\textwidth]{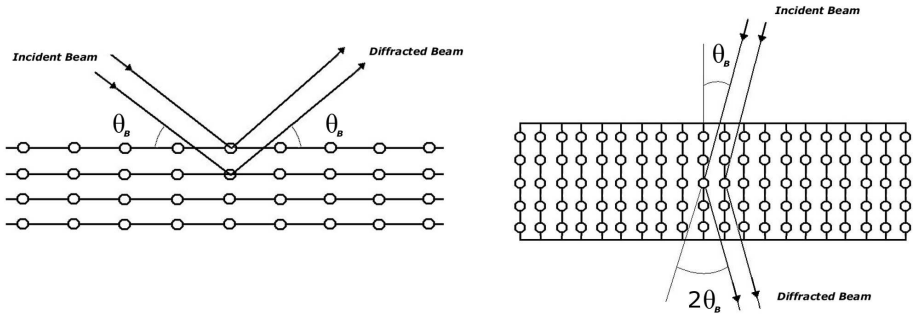}
\caption{The Bragg condition for constructive interference of a gamma-ray photon beam
with the atoms of a given crystalline plane. {\em Left panel:} Bragg diffraction in reflection
configuration (Bragg geometry). {\em Right panel:} Bragg diffraction in transmission 
configuration (Laue geometry).}
\label{f:Bragg}
\end{center}
\end{figure}

\section{Laue lens concept}

Diffraction lenses use the interference\index{Interference} between the periodic nature of
the electromagnetic radiation and a~periodic structure such as the matter in a~crystal. 
For a classical textbook on X--ray diffraction see, e.g., Ref.~\citenum{Zac45}. 
In a Laue lens, the photons 
pass through the full crystal, using its entire volume for interacting coherently. 
In order to be diffracted, an incoming gamma-ray must satisfy the Bragg-condition, 
relating the spacing of lattice planes $d_{hkl}$ with the energy of incident 
photons $E$ and with the angle of incidence $\theta_B$ with respect to the 
chosen set of planes $(hkl)$
\footnote{The indices $h$, $k$, $l$, known 
as {\em Miller indices}, are defined as the reciprocals  of the fractional intercepts 
which the lattice plane makes with the crystallographic axes. For example, if the Miller 
indices of a plane are ({\em hkl}), written
in parentheses, then the plane makes fractional intercepts of $1/h$, $1/k$, $1/l$
with the axes, and, if the axial lengths of the unit cell are $a$, $b$, $c$, the plane makes actual
intercepts of $a/h$, $b/k$, $c/l$. If a plane is parallel  to a given axis, its fractional 
intercept on that axis is taken as infinity and the corresponding Miller index is zero.
If the Miller indices [$hkl$] are shown in square brackets, they give the direction of the
plane with the same indices.}:
%
%
\begin{equation}
2 d_{hkl} \sin \theta_B  = n\frac{hc}{E}
\label{e:bragg}
\end{equation}

where $d_{hkl}$ (in \AA) is the spacing of the lattice planes $(hkl)$, $n$ is 
the diffraction order, $hc = 12.4$~keV$\cdot$\AA\ and $E$ is the energy (in keV) of the gamma-ray
photon.
An elementary illustration
of the Bragg condition, in two different configurations (reflection and transmission), is given in
Fig.~\ref{f:Bragg}, where it can be seen that the incident waves are
reflected by the parallel planes of the atoms in the crystal.

%
%
\begin{figure}[!t]
	\begin{center}
	\includegraphics[width=0.5\textwidth]{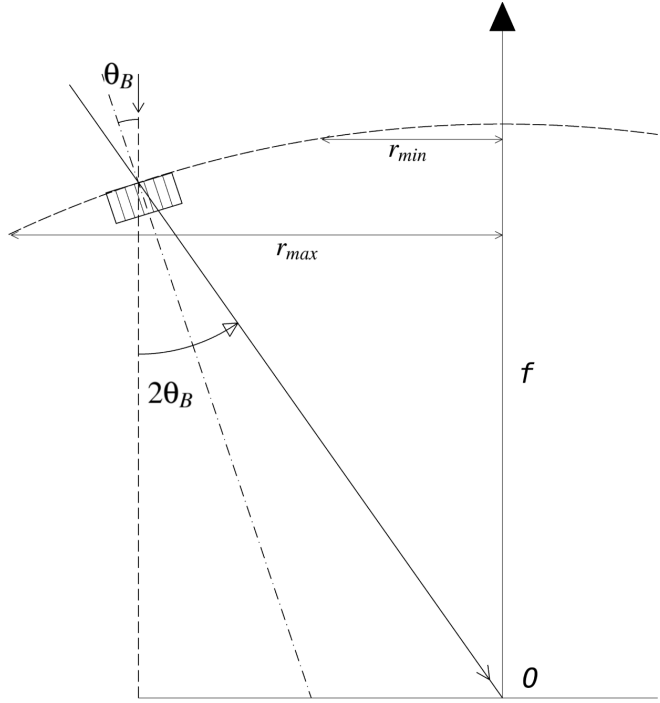}
		\caption{Geometry of a Laue lens (see text).}
		\label{f:CrysPos}
	\end{center}
\end{figure}

A Laue lens  is made of a large number of crystals, in transmission configuration (Laue geometry), 
that are disposed such that they will concentrate the incident radiation onto a common focal spot. 
A convenient way to 
visualize the geometry of a crystal lens is to consider it as a spherical cup 
covered with crystal tiles having their diffracting planes perpendicular 
to the sphere (see Fig.~\ref{f:CrysPos}). The focal spot is on the 
symmetry axis at a distance $f = R/2$ from the cup,  with $R$ being the 
radius of the sphere of which the
spherical  cup is a part;  $f$ is called the {\em focal length}.  

From the Bragg equation, for the first diffraction order ($n = 1$),
it can be seen that the photons incident on a given crystal at 
distance $r$ ($r_{min} \le r \le r_{max}$) from the lens axis can be reflected toward 
the lens focus if their energy $E$ is given by
%
%
\begin{equation}
E = \frac{hc}{2d_{hkl}} \sin \left[ \frac{1}{2} \arctan \left( \frac{f}{r} \right) \right] 
\approx \frac{hc\, f}{d_{hkl}~r}
\label{e:lens-energy}
\end{equation}
where the approximated expression is valid for gamma--ray lenses, given the small 
diffration angles involved.

Conversely, the lens radius $r$ (see Fig.~\ref{f:CrysPos})
at which the photon energy $E$ is reflected in the focus is given
by 
%
%
\begin{equation}
  \mathrm{r} = \mathrm{f}\tan[2\theta_{\mathrm{B}}] 
  \approx \frac{hc f}{d_{hkl} E} 
\label{e:lens-radius}
\end{equation}

\begin{figure}
\begin{center}
\includegraphics[angle=0, width=0.4\textwidth]{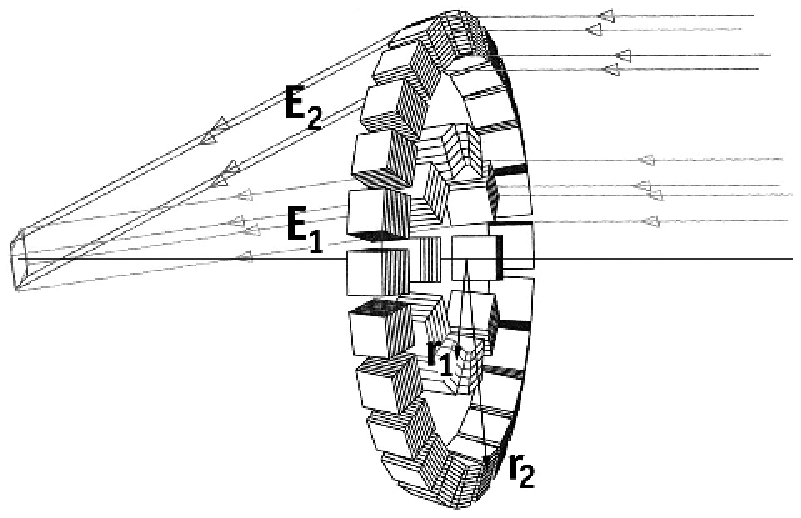}
\includegraphics[width=0.4\textwidth]{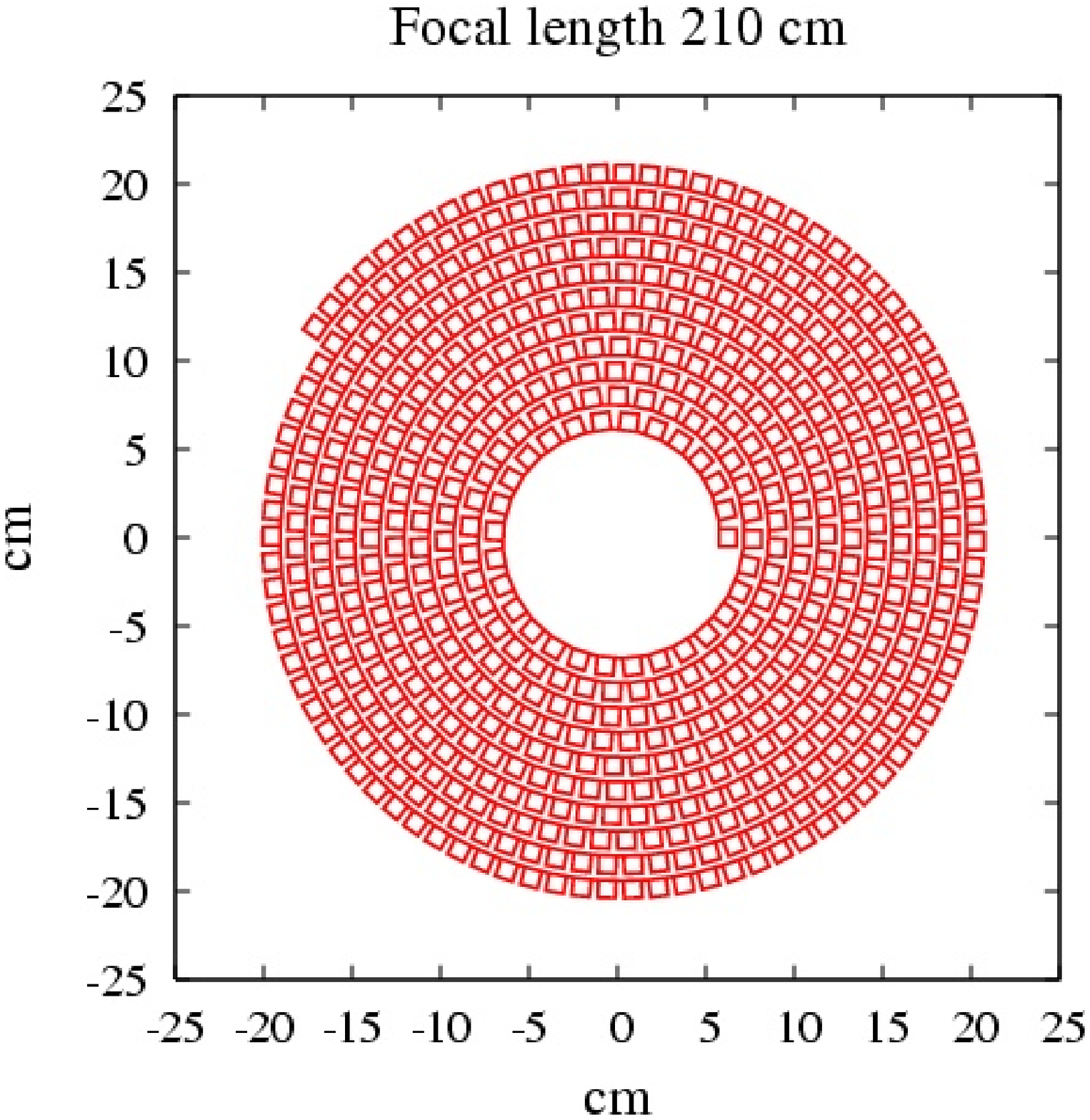}
\end{center}
\caption{The basic design of a crystal diffraction lens in Laue geometry. Flat crystal tiles
	are assumed. {\em Left}: 
	concentric rings of a given radius  $r$ concentrating a constant energy $E$. 
	{\em Right}:  crystal tiles disposed along an Archimedes' spiral result in a 
	continuously varying energy $E$. Given the footprint of the crystals, the image in the 
	focal plane has a minimum size equal to that of the crystal size.}
\label{f:LaueLensPrinciple}
\end{figure}

\begin{figure}[!tb]
	\begin{center}
		\includegraphics[angle=-90,width= 0.5\textwidth]{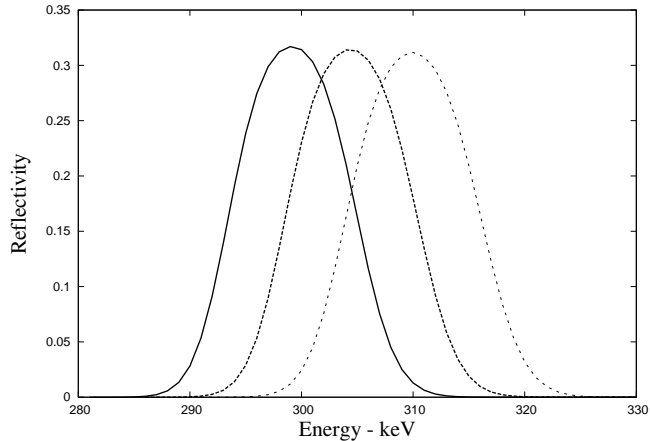}
		\caption{An example of the expected reflectivity profile of three contiguous 
		crystals with a mosaicity of 1.5 arcmin along an Archimedes' spiral. 
		Reprinted from Ref.~\citenum{Pellicciotta06}.} 
		\label{f:refl_spiral}
	\end{center}
\end{figure}

Rotation around the lens optical axis at constant $r$
results in concentric rings of crystals (see left panel of Fig.~\ref{f:LaueLensPrinciple}), while a 
uniformly changing value of $r$ gives rise to an Archimedes spiral (right panel of 
Fig.~\ref{f:LaueLensPrinciple}).  Assuming that the chosen diffracting planes 
($hkl$) of all the lens crystals are the same,
in the first case (constant $r$) the energy of the diffracted photon will be centered on~$E$ for 
all the crystals in the ring, while in the second case (Archimedes spiral), 
the reflected energy $E$ will continuously vary from one crystal to the other, as shown in 
Fig.~\ref{f:refl_spiral}.
%

%
%

%
%

\subsection{Energy passband}

Any Laue lens will diffract photons over a certain 
energy passband ($E_{min}$,$E_{max}$). From eq.~\ref{e:lens-energy}, at first order diffraction 
(the most efficient), it results that

%
%
\begin{equation}
	E_{min} \approx \frac{hc f}{d_{hkl}~r_{max}}
\label{e:Emin}
\end{equation}
%
%
%
\begin{equation}
	E_{max} \approx \frac{hc f}{d_{hkl}~r_{min}}.
\label{e:Emax}
\end{equation}

Given that, for astronomical applications, the lens passband is desired to be covered with 
the highest effective area 
\footnote{The effective area at energy $E$ is defined as the geometrical area of the lens 
projected in the focal plane times 
the total reflection efficiency at energy $E$.} and in a smooth manner as a function of energy, 
the energy bands of the 
photons reflected by contiguous crystal rings or, in the case of the Archimedes structure 
of a lens, by contiguous crystals, have to overlap each other, like in Fig.~\ref{f:refl_spiral}. 
Since the full width at half maximum (fwhm) of the acceptance angle  $\delta$  
(known as {\em the Darwin width}) of perfect crystals is extremely narrow 
(fractions of an arcsec to a few arcsec, see Ref.~\citenum{Zac45}), such materials 
are not suitable for astrophysical Laue lenses.
In order to increase the energy passband of individual crystals one uses mosaic 
crystals or curved crystals (see following section). The {\em mosaicity} of mosaic crystals  (see 
Sect.~\ref{s:mosaic_cryst}) and the {\em total bending angle} of curved crystals (see 
Sect.~\ref{s:bent_cryst})  govern
the flux throughput, the angular resolution and the energy passband  of the Laue lenses.  
The diffracted flux from a~continuum source increases with
increasing the mosaicity of mosaic crystals or the total bending angle of curved crystals. 
For a~crystal lens telescope, crystals with mosaicities or total bending angles ranging 
from a~few tens of arcseconds to a~few arcminutes are of interest.

The bandwidth of a lens for an on--axis source is determined by the mosaicity or total bending angle
of the individual crystals and by the accuracy of the alignment of the single
crystals. By forming the derivative of the Bragg relation\index{Bragg relation} in the small
angle approximation ($2 d_{hkl}\theta_B \approx n h c/E$), we get
\begin{equation}
  \label{eq:4-24}
  \Delta \theta_B/\theta_B = \Delta E/E \;.
\end{equation}
If $\Delta \theta_B$ is the mosaicity of the mosaic crystal or the total bending angle of the curved crystal,
the corresponding energy passband $\Delta E $ of the crystal becomes
\begin{equation}
  \label{e:DeltaE}
  \Delta E = \frac{2 d_{hkl}\cdot E^2 \cdot \Delta \theta_B}{n h c}\;.
\end{equation}
It is worth pointing out that, whereas the energy passband of a~crystal lens grows with the 
square of energy, the Doppler broadening of the astrophysical lines (e.g. in SN ejecta)
increases linearly with energy for a~given expansion velocity.

\section{Crystal reflectivity}

Both mosaic crystals and curved crystals are suitable to be used for a Laue lens. We 
discuss the properties of both of them and their reflectivity. 

\subsection{Mosaic crystals}
\label{s:mosaic_cryst}

Mosaic crystals are made of many microscopic perfect crystals (\textit{crystallites}) 
with their lattice planes slightly misaligned with each other around a mean direction, 
corresponding to the mean lattice planes $(hkl)$ chosen for diffraction.
In the lens configuration assumed, the mean lattice plane is normal to the surface of
the crystals. The distribution function of the crystallite misalignments
from the mean direction can be approximated by a Gaussian function:
\begin{equation}
\label{eq:Gauss}
W(\Delta)=\frac{1}{\sqrt{2\pi}\eta}
\exp{\left( - \frac{\Delta^2}{2\eta^2}\right )} \, ,
\end{equation}
where $\Delta$ is the magnitude of the angular deviation from the mean, while 
$\beta_m = 2.35 \eta$ is the fwhm of the mosaic spread (called {\em mosaicity}). 

For the Laue geometry and diffracting planes perpendicular to the cross section 
of the crystal tile (see, e. g., Fig.~\ref{f:CrysPos}),
the crystal reflectivity $R(\Delta, E)$ is  given 
by Ref.~\citenum{Zac45}:
\begin{equation}
R(\Delta, E) = \frac {I_d (\Delta, E)}{I_0} = \sinh{(\sigma T)}
\exp{\left [ - \left (\mu + \gamma _0 \sigma \right )
\frac {T}{\gamma _0} \right ]} = 
\frac{1}{2}(1- e^{-2\sigma T}) e^{ - \mu \frac {T}{\gamma _0}}
\, ,
\label{e:refl_mosaic}
\end{equation}
where $I_0$ is the intensity of the incident beam, $\mu$ is the absorption 
coefficient corresponding to that energy, $\gamma_0$ is the cosine of
the angle between the direction of the photons and the normal to 
the crystal surface, $T$ is the thickness of the mosaic crystal and $\sigma$ is: 
\begin{equation}
\label{eq:sigma}
\sigma = \sigma (E, \Delta) = W(\Delta) Q(E) f(A) \, ,
\end{equation}
where
\begin{equation}
Q (E) =  r_e^2 \, \left | \frac {F_{hkl}}{V}\right |^2 \,
\lambda ^3 \, \frac {1+\cos^2(2 \theta_B)}{2 \sin 2\theta_B} \, ,
\label{eq:q}
\end{equation}
in which $r_e$ is the classical electron radius, $F_{hkl}$ is the structure
factor, inclusive of the temperature effect (Debye-Waller's factor), 
$V$ is the volume of the crystal unit cell, $\lambda$ is the radiation wavelength 
and $\theta_B$ is the Bragg angle for that particular energy, 
while $f(A)$ in Eq.~\ref{eq:sigma} is well approximated by:
%
%
\begin{equation}
f(A)= \frac{B_0(2A) + |\cos2\theta_B| \,  B_0(2A|\cos2\theta_B|)}{2A(1+ \cos^2\theta_B)}. \\
\end{equation}
Here $B_0$ is the Bessel function of zero order integrated between 0 and $2A$, with 
$A$ defined as follows:
 \begin{equation}
 A = \frac{\pi \, t_0}{\Lambda_0 \, \cos \theta_B},
 \end{equation}
in which $t_0$ is the crystallite thickness, and $\Lambda_0$ ({\em extinction 
length}) is defined for the symmetrical Laue case 
(see e.g. ref.~\citenum{Authier01}) as:
\begin{equation}
\Lambda_0 = \frac{\pi\, V \, \cos \theta_B}{r_e \lambda \, |F_{hkl}| \, (1+ |\cos 2\theta_B|)},
\label{e:extin_length}
\end{equation} 
In general $f(A)<1$ and converges to 1 if $t_0 \ll \Lambda_0$. In this 
case we get the highest reflectivity.

The quantity $\gamma_0\sigma$ is known as {\em secondary extinction} 
coefficient and $T/\gamma_0$ is the 
distance travelled by the direct beam inside the crystal.

\subsection{Curved crystals}
\label{s:bent_cryst}

Similarly to mosaic crystals, curved crystals have an angular dispersion of
the lattice planes and thus a much larger energy passband (see Eq.~\ref{e:DeltaE})
than perfect crystals. The properties of these crystals and the methods to get
them are discussed
in Ref.~\citenum{Barriere09}. Here we summarize their reflection
properties.

The most recent theory of the radiation diffraction in transmission
geometry for such crystals, in the case of a large and homogeneous curvature, is now
well fixed and has been compared with the experimental results (see 
Refs.~\citenum{Keitel99,Malgrange02}). 
In this theory, the distortion of diffracting planes is described 
by the strain gradient $\beta_s$, that, in the case of a uniform curvature, is given
by:
\begin{equation}
\beta_s = \frac{\Omega}{T_0 (\delta/2)}
\end{equation}
where $\Omega$ is the {\em total bending angle} and corresponds to the mosaicity of the 
mosaic crystals, $T_0$ is the thickness of the crystal and $\delta$ is the Darwin width. 

\begin{figure}
\begin{center}
\includegraphics[angle=0, width=0.4\textwidth]{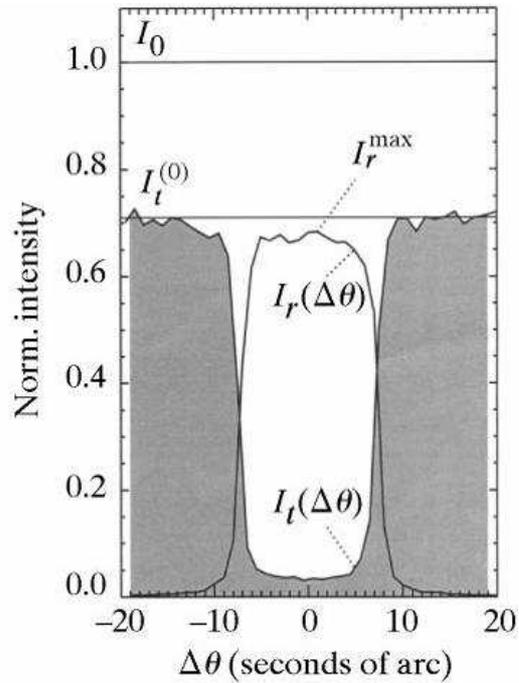}
\end{center}
\caption{Reflectivity profile of a curved crystal as a function of the rocking angle 
$\Delta \theta$. This angle gives the difference between the incidence angle of the monochromatic 
photon beam and the Bragg angle. $I_0$ represents the incident intensity, $I_t(\Delta \theta)$ 
the transmitted intensity and $I_r(\Delta \theta)$ the reflected intensity. Reprinted 
from Ref.~\citenum{Keitel99}.}
\label{f:refl_curved}
\end{figure}

When the strain gradient $|\beta_s|$ becomes larger than a critical value 
$\beta_c = \pi /(2 \Lambda_0)$, it has been shown that, for
a uniform curvature of planes, the peak reflectivity $R^{max}$ of a curved crystal 
is given by:
\begin{equation}
R^{peak}(c_p, E) = \frac{I^{peak}_r(c_p, E)}{I_0}  = \left( 1 - e^{-{\pi^2 d_{hkl} \over c_p \, \Lambda_0^2}} \right) \, e^{-{\mu \, \Omega \over c_p \, \cos \theta_B}}.
\label{e:refl_curved}
\end{equation}
where $I_r^{peak}$ is the reflected peak intensity, $c_p = \Omega/T_0$ is 
the curvature of the lattice planes assumed to be uniform
across the crystal thickness, and the extinction length $\Lambda_0 = \Lambda_0(E)$ 
is given by Eq.~\ref{e:extin_length}. The reflected intensity profile $I_r(c_p, E)$
is that of a perfect crystal with the Darwin width replaced with $\Omega$. This profile
is shown in Fig.~\ref{f:refl_curved}. 
%
%

From the last equation, it can be shown that the highest peak reflectivity is obtained
for a curvature of the lattice planes given by:
\begin{equation}
c_p^{opt} = \frac{M} {\ln \left( 1 + {M \over N} \right) }.
\end{equation}
where $M = \frac{\pi^2 d_{hkl}}{\Lambda_0^2}$ and  $N= \frac{\mu \Omega}{\cos \theta_B}$.

\subsection{Mosaic crystals vs. curved crystals}
\label{s:curved}

Both mosaic crystals and curved crystals can be used for a Laue lens, if
they can be produced with the needed angular spread.
However, in principle, curved crystals can reach a higher efficiency  than mosaic
crystals. Indeed, while the diffraction efficiency of mosaic crystals
is limited to 50\%, that of curved crystals can reach 100\%. 
Another advantage of curved crystals is that the diffraction profile
of a curved crystal is rectangular with width related to $\Omega$,
while that of mosaic crystals is Gaussian with fwhm equal to the
mosaicity $\beta_m$. Given the absence of Gaussian tails, curved crystals 
concentrate the flux better (see next Section). This better performance
of the curved crystals with respect to the mosaic crystals for Laue
lenses is discussed in depth in Ref.~\citenum{Barriere08}. 

\begin{figure}
\begin{center}
\includegraphics[angle=0, width=0.4\textwidth]{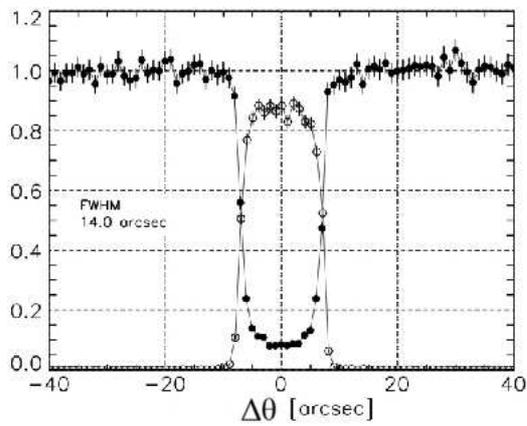}
\end{center}
\caption{Measured  rocking curve, in transmission geometry at 150 keV, of a Si(111) crystal  
curved at the University of Ferrara (see text). $\Delta \theta$ gives the difference between the 
incidence angle of the monochromatic photon beam and the Bragg angle. {\em Open circles}: ratio between measured 
intensity of the diffracted beam and measured intensity of the transmitted beam 
(also called {\em diffraction efficiency}). Filled circles: difference between 
transmitted and diffracted intensities. Note that the angle $\Delta \theta$, 
through the Bragg law, is related to the reflected photon energy. Thus the figure also
shows the energy bandwidth of the curved crystal. 
Reprinted from Ref.~\citenum{Barriere09c}, who tested the crystal sample.}
\label{f:Si_Guidi}
\end{figure}

Curved crystals can be obtained in various ways \cite{Barriere09}. 
The most feasible techniques to
be used for Laue lenses include the elastic bending of a perfect 
crystal (the technique commonly adopted in synchrotron radiation facilities), 
the deposition of a coating on a wafer, the growing of a two-component crystal
whose composition varies along the crystal growth axis (see, e.g., Ref.~\citenum{Keitel99}) or 
the indentation of one face of a wafer. The last technique is being developed at the University
of Ferrara (V. Guidi, private communication) with very satisfactory 
results (see Fig.~\ref{f:Si_Guidi}). Also the deposition of a coating on a wafer is being
tested at the same University.
%
%

\section{Optimization of a Laue lens}

The free parameters of a Laue lens are the crystal properties (materials, lattice planes
for diffraction, micro-crystal size and mosaicity in the case of mosaic crystals, 
total bending angle in the case of curved crystals, crystal thickness), the lens geometry 
(ring-like or Archimedes' spiral), 
its focal length and its nominal energy passband.
Many optimization studies of these parameters have been performed and tested 
\cite{Pisa04,Pisa05a,Pisa05b,Barriere09}.

\subsection{Crystal material}

\begin{figure}
\begin{center}
\includegraphics[angle=0, width=0.4\textwidth]{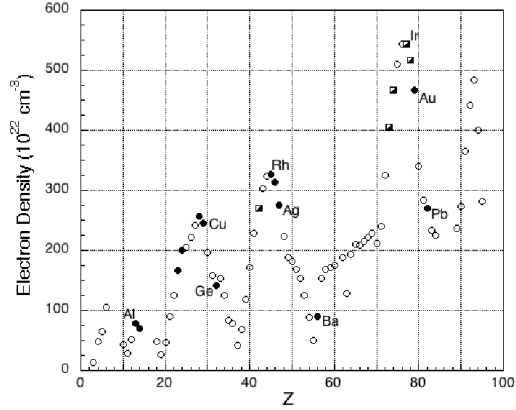}
\end{center}
\caption{Density of a crystal unit cell versus element atomic number. Reprinted 
from Ref.~\citenum{Barriere09}.}
\label{f:N}
\end{figure}

Independently of the crystal structure (mosaic or curved), in order to optimize the
crystal reflectivity it is important to maximize $Q(E)$ as defined in Eq.~\ref{eq:q}. 
This is the integrated crystal reflectivity per unit volume, whose normalization 
is the ratio $|F_{hkl}/V|^2$ between the structure factor of the chosen lattice
planes $F_{hkl}$ and the volume $V$ of the unit cell. The inverse of $V$ is the
atomic density $N$. Thus a small $V$ (or a high $N$) is important for maximizing $Q(E)$.
The value of $N$ as a function the element atomic
number $Z$ is shown in Fig.~\ref{f:N}.
%
%

\begin{figure}
\begin{center}
\includegraphics[angle=-90,width=0.6\textwidth]{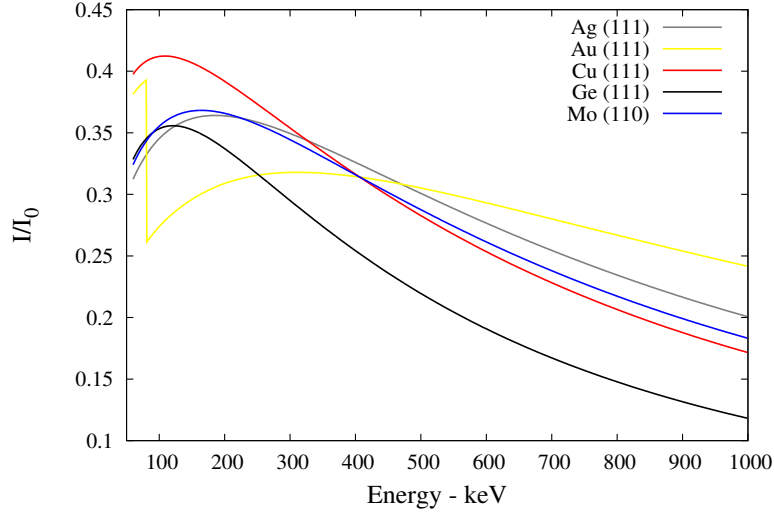}
\caption{Peak reflectivity of 5 candidate crystal materials. The Miller indices 
used give the highest reflectivity. A 
mosaicity of 40 arcsec is assumed. The thickness has been optimized.
The production technology of
mosaic crystals with the required spread is already mature for Ge and Cu.}
\label{f:materials}
\end{center}  
\end{figure}

As can be seen, for single--element materials, broad density peaks are apparent 
in correspondence of the atomic numbers
5, 13, 28, 45, and 78. Common materials like Al ($Z=13$), Si ($Z=14$), 
Ni ($Z=28$), Cu ($Z=29$),  Zn ($Z=30$), Ge ($Z=32$), Mo ($Z=42$), Rh ($Z= 45$), Ag ($Z=47$), 
Ta ($Z=73$), W ($Z=74$), Au ($Z=42$) are good candidates to be used for Laue lenses and should 
be preferred to other elements if they are available as crystals with the requested properties.
The peak reflectivity versus energy of few single--element mosaic crystal materials is shown in 
Fig.~\ref{f:materials}.

%
%

Also double--element crystal materials can be used for Laue lenses. 
Several of them, developed for other applications, are already available, 
like GaAs, InAs, CdTe, CaF$_2$. With some improvements, these crystal materials
can be used for Laue lenses (see discussion in Ref.~\citenum{Barriere09}).

Clearly the best lattice planes are those that optimize the structure factor $|F_{hkl}|$, 
under the condition that the corresponding $d_{hkl}$ is consistent with lens 
constraints, such as energy passband, lens size and focal length (see below).
  
\subsection{Crystallite size and angular distribution in mosaic crystals}

From the reflectivity equation (Eq.~\ref{e:refl_mosaic}), the 
crystallite thickness $t_0$ plays an important role in the reflectivity  optimization. 
For fixed values
of the mosaicity and crystal thickness, the highest reflectivity
is obtained for a crystallite thickness that satisfies the 
condition $t_0 \ll \Lambda_0$. In general, this implies a thickness of the order of 1~$\mu$m. 

Unfortunately this condition is still not always satisfied.
From extended tests performed on Cu(111) supplied by ILL \cite{Courtois05}, it was found that 
the condition above
is satisfied in single points \cite{Pellicciotta06}, but not when 
the entire crystal cross section is irradiated (values even higher than 100~$\mu$m have been 
found \cite{Barriere09}).
In  addition,  in  Ref.~\citenum{Barriere09} it is found that $t_0$ is energy dependent, which
is a surprising result that requires an interpretation (see discussion therein).

\begin{figure}
\begin{center}
\includegraphics[width=0.45\textwidth]{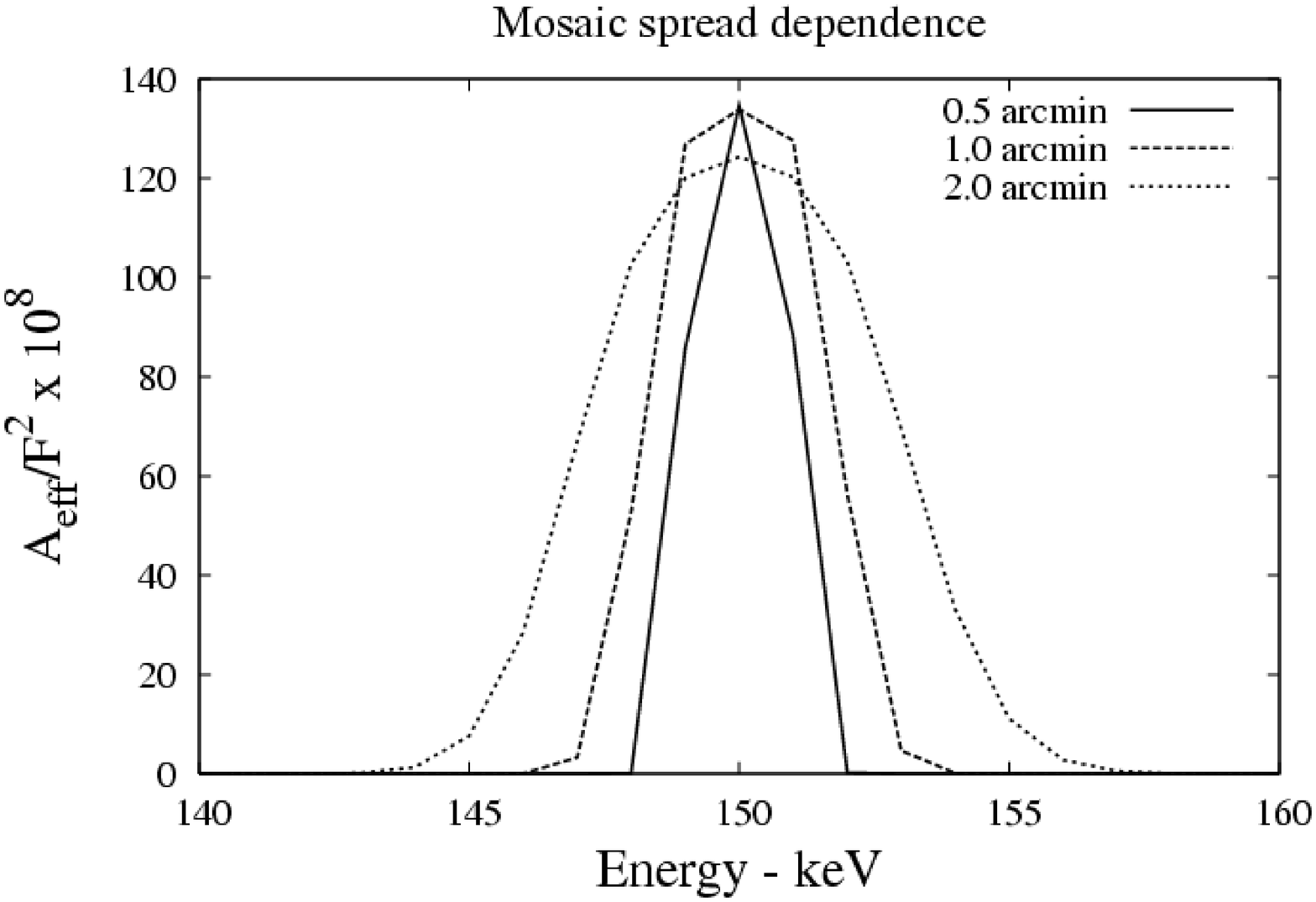}
\includegraphics[width=0.45\textwidth]{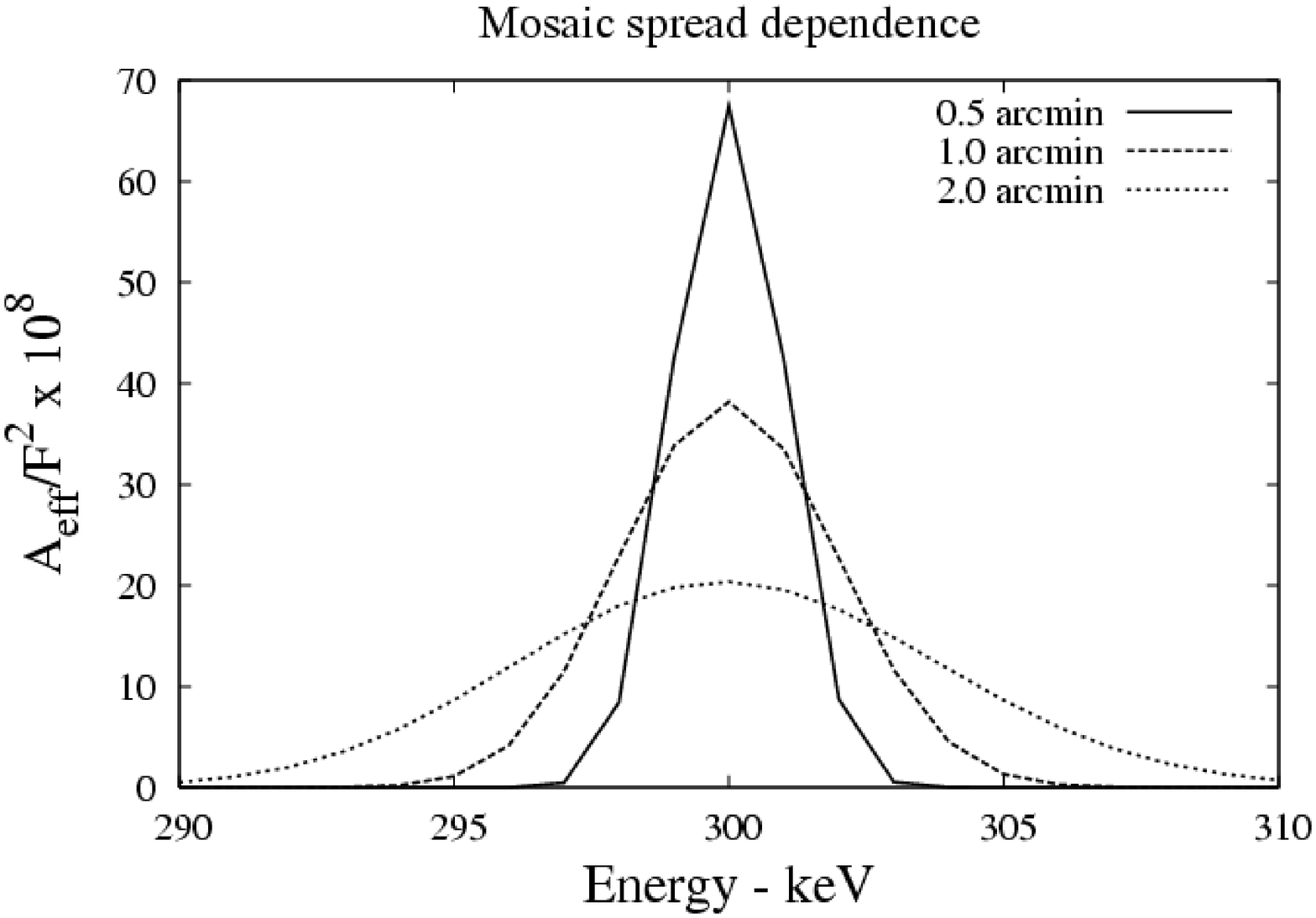}
\end{center}
\caption{Normalized effective area for different values of the mosaicity of Cu(111). 
\textit{Left}: First diffraction order; \textit{Right}: second diffraction order.
Reprinted from Ref.~\citenum{Pisa05b}.}
\label{f:spreadarea}
\end{figure}

The crystal mosaicity is another crucial parameter for the optimization 
of the lens performance. 
It can be seen \cite{Pisa04,Pisa05a,Pisa05b,Frontera06} that, even if a higher mosaicity 
gives a larger lens effective area (see Fig.~\ref{f:spreadarea}), a higher spread  
also produces a larger defocusing of the reflected photons in the focal plane and thus
a lower lens sensitivity. 
 
%
%
This can be seen by introducing the focusing factor $G (E)$ of a Laue lens:
\begin{equation}
G(E)= f_{ph} \frac{A_{eff}(E)}{A_d}
\end{equation}
in which $A_{eff}(E)$ is the effective area of the lens and $A_d$ is the
area of the focal spot which contains a fraction $f_{ph}$ of photons reflected by the lens.
 
%
%

\begin{figure}
\begin{center}
\includegraphics[angle=270,width=0.49\textwidth]{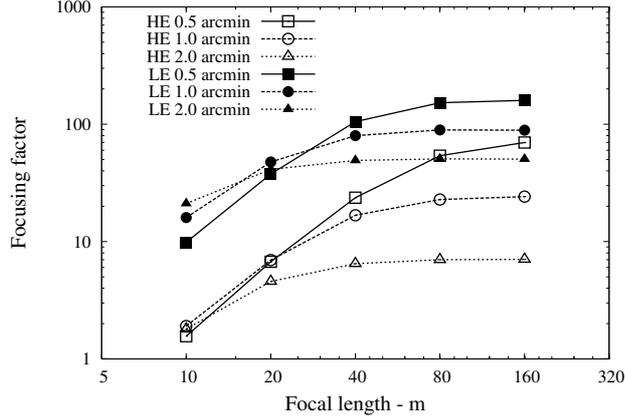}
\caption{Dependence of the focusing factor $G$ of a lens made of mosaic crystals of Cu(111)
 on the focal length $f$ for three different values of the mosaicity in two different energy 
bands: 90--110 keV (LE) and 450--550 keV (HE). Reprinted from Ref.~\citenum{Frontera06}.}
\label{f:G_vs_F_and_spread}
\end{center}
\end{figure}

Assuming that the detector noise is Poissonian, it can be easily shown that $G(E)$ is inversely proportional 
to the minimum detectable continuum intensity of a lens: 
\begin{equation}
I_{min} (E) = \frac{n_\sigma}{\eta_d \, G} \sqrt{\frac{2 B}{A_d \, \Delta T \, \Delta E}}
\label{e:sens}
\end{equation}
where $I_{min}$ is given in units of photons~cm$^{-2}$~s$^{-1}$~keV$^{-1}$ 
in the interval $\Delta E$  around $E$, $n_\sigma$ is the significance
level of the signal (typically $n_\sigma = $3--5), $B$ is the focal plane detector
background intensity (counts~cm$^{-2}$~s$^{-1}$~keV$^{-1}$), $\Delta T$ is exposure time to 
a celestial source, and $\eta_d$ is the focal plane detector efficiency 
at energy $E$. For a lens made of mosaic crystals of Cu(111), Fig.~\ref{f:G_vs_F_and_spread} shows the 
dependence of $G$ on focal length 
in two different energy bands, for different values of the mosaicity.
%
%

As can be seen, in spite of the fact that a higher spread gives a higher effective area,
a lower spread gives a higher $G$ and thus a higher lens sensitivity. This
effect is small at short focal lengths (10--20 m), but it becomes very significant at long focal
lengths ($>$30~m), especially at high energies ($\sim$ 500~keV).

\subsection{Crystal thickness}

\begin{figure}
\begin{center}
\includegraphics[angle=270,width=0.49\textwidth]{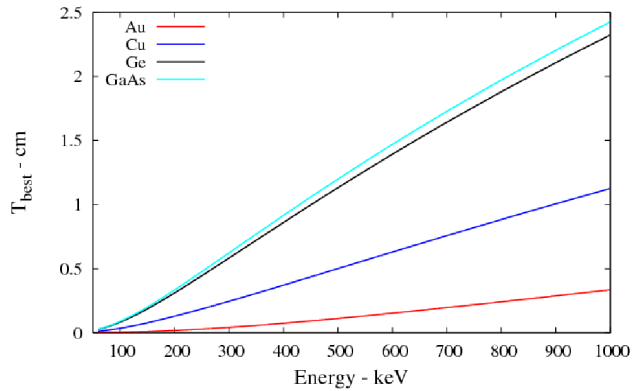}
\caption{The best crystal thickness that maximizes the crystal reflectivity, for various
materials. The mosaicity assumed is 1 arcmin, the crystallite thickness
is 1~$\mu$m, and the crystal plane chosen for all of them is (111).}
\label{f:thickness}
\end{center}
\end{figure}

The crystal thickness is another crucial parameter for the reflectivity optimization of 
a mosaic crystal. The best crystal thickness  is given by

\begin{equation}
T_{best} = \frac 1 {2 \sigma} 
\ln {\left ( 1 + \frac {2\sigma \gamma _0}{\mu} \right )} \, 
\label{eq:tmbest}
\end{equation}
in the case of a mosaic crystal, and it is given by
\begin{equation}
T_{best} = {\Omega \,  \ln \left( 1 + {M \over N} \right) \over M}.
\label{e:tcbest}
\end{equation}
in the case of a curved crystal, where the involved quantities are defined in the sections
above.
In the case of a mosaic crystal, the optimum crystal thickness for various materials
is shown in Fig.~\ref{f:thickness}.
%
%

As can be seen, the best crystal thickness depends on the absorption coefficient $\mu$. 
A high absorption coefficient implies a low crystal thickness for the reflectivity optimization.

\subsection{Focal length}

The focal length has a key importance in the case of Laue lenses,  more than its importance
in the case of traditional focusing telescopes. Indeed, given that the energy 
passband of a single crystal is quite narrow (see e.g., Fig.~\ref{f:refl_spiral}), 
from the expression of nominal energy passband ($E_{min}$, $E_{max}$) of a lens
(see eqs.~\ref{e:Emin} and \ref{e:Emax}), it results that,  
for first order diffraction, the lens radii ($r_{min}$, $r_{max}$), needed to get 
a given passband, depend linearly on $f$:
%
%
\begin{equation}
	r_{min} \approx \frac{hc f}{d_{hkl}~E_{max}}
\end{equation}
%
%
%
\begin{equation}
	r_{max} \approx \frac{hc f}{d_{hkl}~E_{min}}.
\end{equation}
Given that high energy photons are focused by the innermost part of the lens,
to get a given lens inner area, the focal length must be increased 
(the lens area approximately increases with $f^2$). 

A~gamma-ray lens with a broad
continuum coverage from 300 keV to 1.5 MeV was proposed in the 90's 
by N.~Lund \cite{Lund92}. He assumed mosaic crystals of Copper and Gold. 
In order to achieve a significant effective area
at high energies ($350\,\mathrm{cm}^2$ at 300\,keV and $25\,\mathrm{cm}^2$ at 1.3\,MeV),
the focal length proposed was 60 m.

\subsection{Broad vs. narrow passband Laue lenses} 

 For the lens optimization, the selection criteria of the crystal material and lattice planes 
can change depending on the requested lens passband.
Two classes of Laue lenses can be identified, \emph{broad} passband  and \emph{narrow} 
passband. The former covers a broad energy band (e.g., 100-600 keV)
for the study of continuum source spectra, while the latter achieves an optimal sensitivity 
in a relatively narrow energy band (e.g., 800-900 keV) for gamma-ray line spectroscopy.
These two classes of lenses require different criteria in the crystal choice and disposition
in the lens for its optimization.

\subsubsection{Narrow passband Laue lenses}

Assuming a ring--like geometry, these lenses use a~different crystalline plane
\emph{(hkl)} for every ring in order to diffract photons in only one
energy band centered on the energy $E_0$.
If, for a given focal length, the ring at distance $r_0$ from the lens axis 
concentrates photons centered at energy $E_0$ using crystals with 
crystalline plane spacing $d_0$, a~ring with a~radius $\mathrm{r}_{1} > \mathrm{r}_{0}$, 
according to  the Bragg law, will concentrate the same 
photon energies only if the crystalline plane spacing~$\mathrm{d}_{1}$ is smaller 
than~$\mathrm{d}_{0}$ or if a~higher order is used. 
From Eq.~\ref{e:lens-radius}, for materials with a~cubic structure for which
$d_{hkl}$ is inversely proportional to $\sqrt{\mathrm{h}^2+\mathrm{k}^2+\mathrm{l}^2}$,
the ring radii are proportional to this quantity. 

As the diffraction efficiency decreases with increasing diffraction
order $n$, a~crystal in an exterior ring will add less efficient area
to the lens than a~crystal on an inner ring. However, since the number
of crystals increases with the ring-radius, all rings will usually
contribute about the same amount of efficient area to the lens. Using
larger and larger Bragg angles with increasing ring radius allows the
instrument to be relatively compact, featuring a shorter focal
length than required if the above criterion were not adopted. 
An example of a~narrow passband Laue lens, the balloon telescope CLAIRE, 
will be discussed below.

\subsubsection{Broad passband Laue lenses}
\label{s:broad_band}
\begin{figure}
\begin{center}
\includegraphics[angle=0,width=0.40\textwidth]{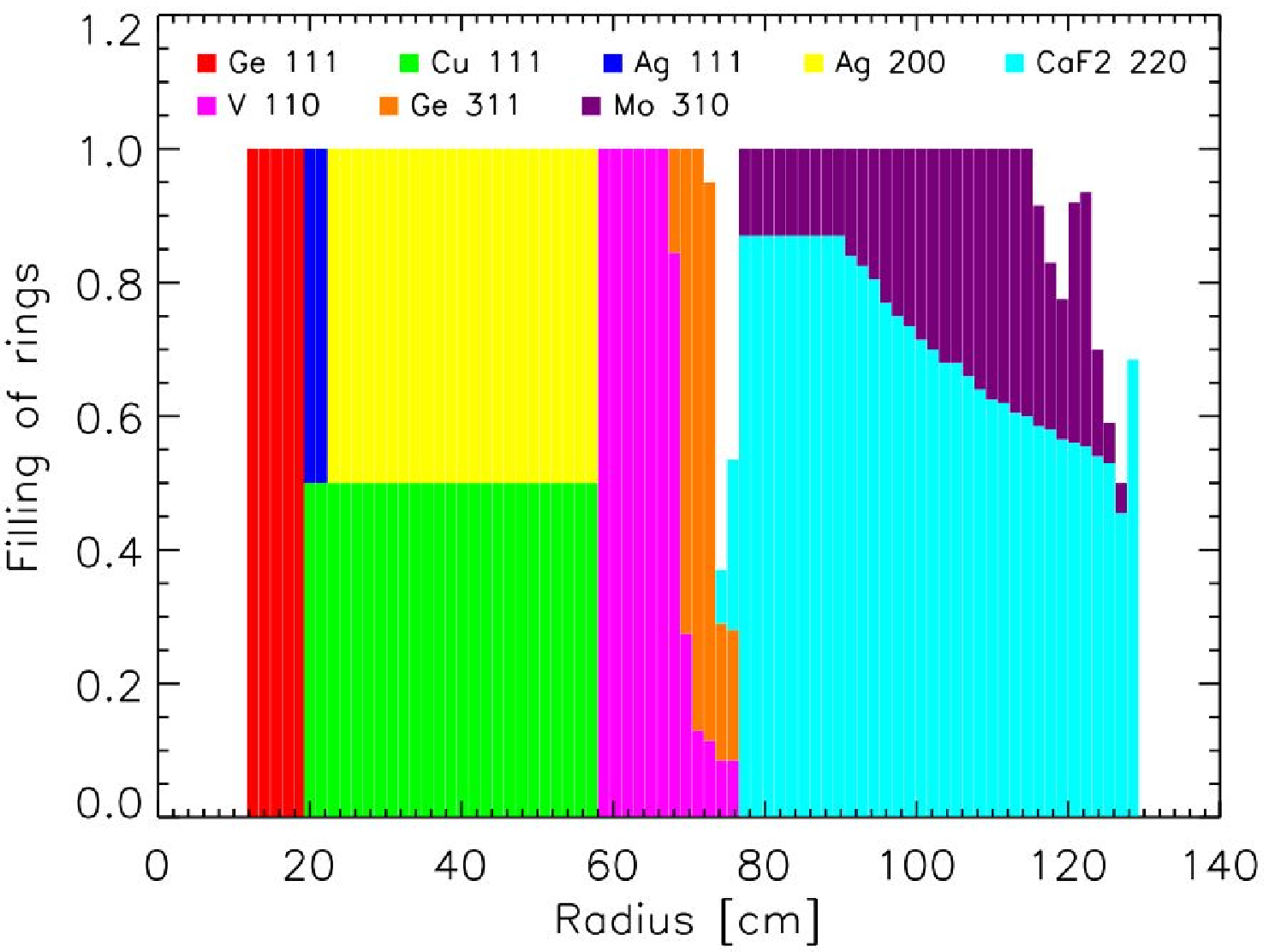}
\includegraphics[angle=90,width=0.4\textwidth]{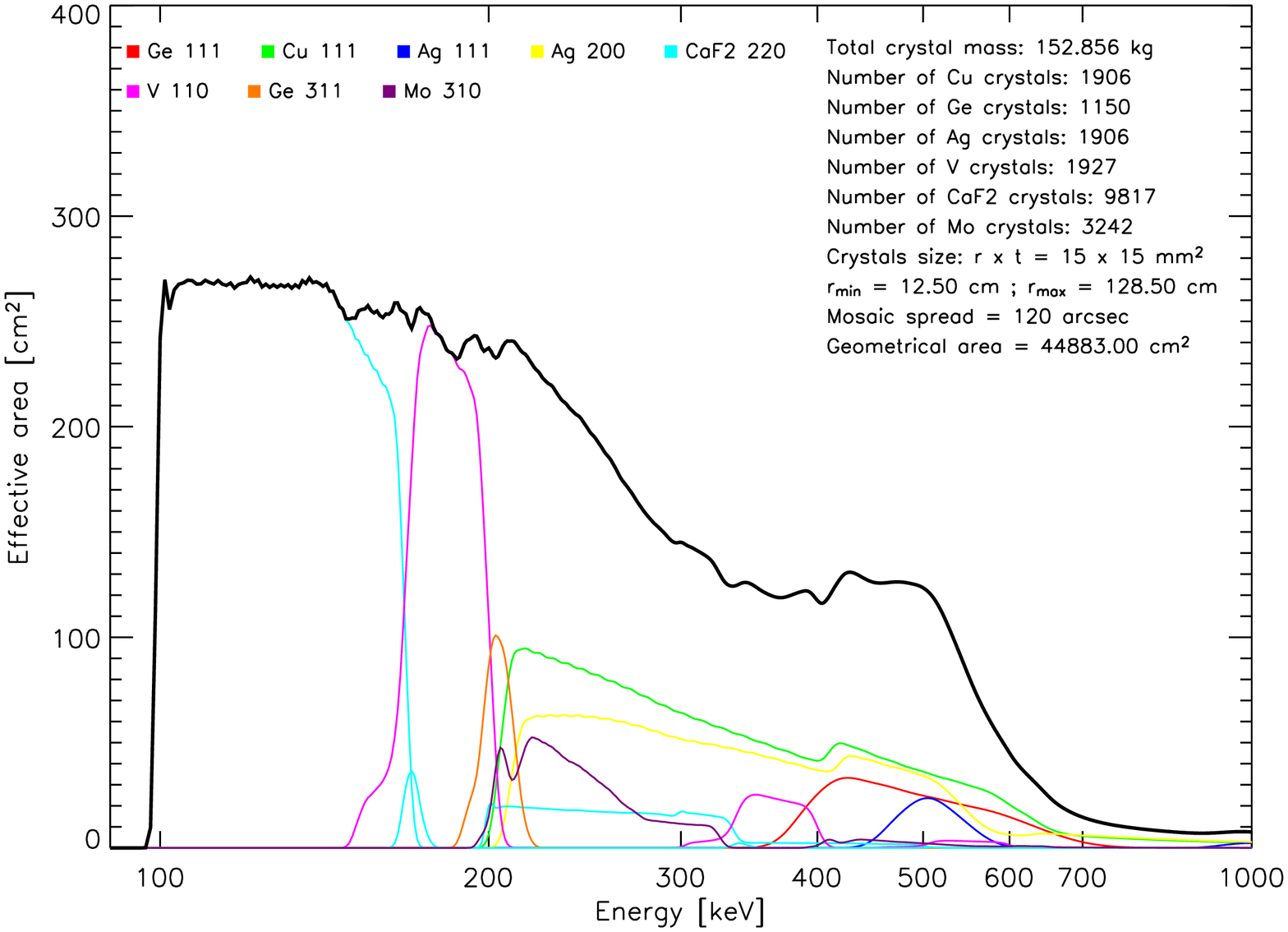}
\hspace{0.3cm}
\caption{{\em Left}: Distribution of various flat mosaic crystals and chosen lattice planes in an
example of a lens and focal length, as studied by Barriere et 
al.~\cite{Barriere09c}. {\em Right}: Effective area of the studied lens. Colors show the contribution 
of each material.} 
\label{f:crystal-distr}
\end{center}
\end{figure}

These lenses use the best combination of crystalline planes to cover
in the most efficent way the lens passband. Lowest order planes, e.g. $(111)$,
are preferred because we can exploit, in addition to their optimum diffraction 
efficiency, also the higher order diffraction of the same planes. 
The principle used for covering a broad energy band is the following.
In the simple assumption that a single crystal material and crystal plane $(hkl)$ are used,
assuming a ring-like geometry of the lens, different concentric rings focus slightly
different energies because of the varying Bragg angle, and thus with with several rings
a~broad energy band can be covered with the first order diffraction (lens nominal 
energy passband) . But, in addition to the first order diffraction, higher order
diffraction can be exploited that increases the effective area at higher energies, while
also extending the passband.
An example of the distribution of the various crystals and lattice planes in different rings
for a broad passband lens (100--600 keV) made of mosaic crystals, with 20 m focal length, 
is studied in Ref.~\citenum{Barriere09c}, and it is shown in Fig.~\ref{f:crystal-distr}.

%
%

Diffraction lenses with broad energy passbands were also developed and
tested for low energy X-rays since the sixties (e.g. Lindquist and
Webber~\cite{Lindquist68}). Today, photons up to 80 keV can be efficiently
focused thanks to the development of grazing incidence based on  multilayer mirrors (see other 
papers in this issue). Above this upper threshold, 
the development of Laue lens telescopes becomes crucial for photon focusing.

\section{Optical properties of a lens}

The optical properties of a lens, for both on--axis and off--axis incident photons, 
have been investigated by means of Monte Carlo (MC) simulations \cite{Pisa05b,Frontera06}.
For flat crystal tiles, the Point Spread Function (PSF) depends on the crystal size, on their mosaicity
and on the accuracy of their positioning in the lens. 
 
\begin{figure}
\begin{center}
\includegraphics[angle=0, width=0.4\textwidth]{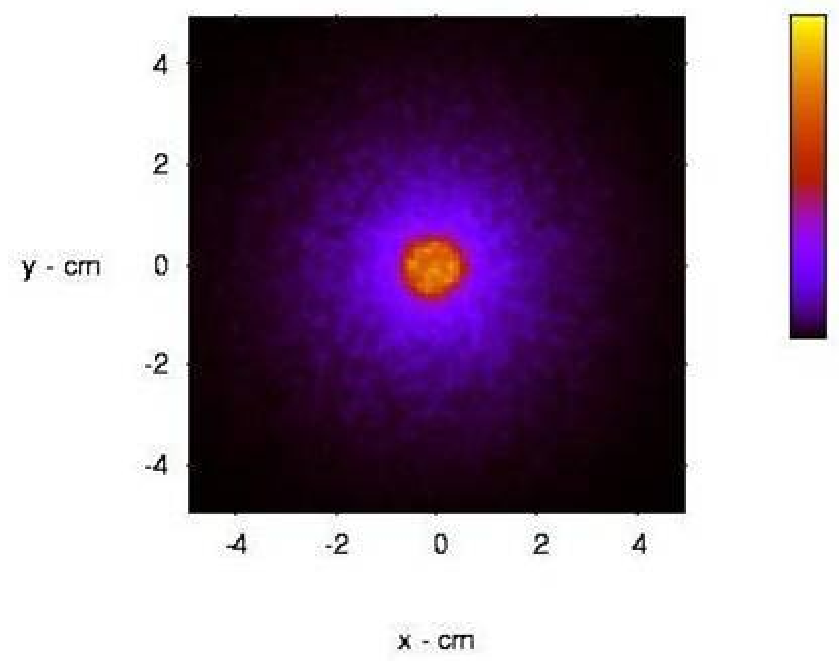}
\raisebox{0.5cm}{\includegraphics[angle=0, width=0.4\textwidth]{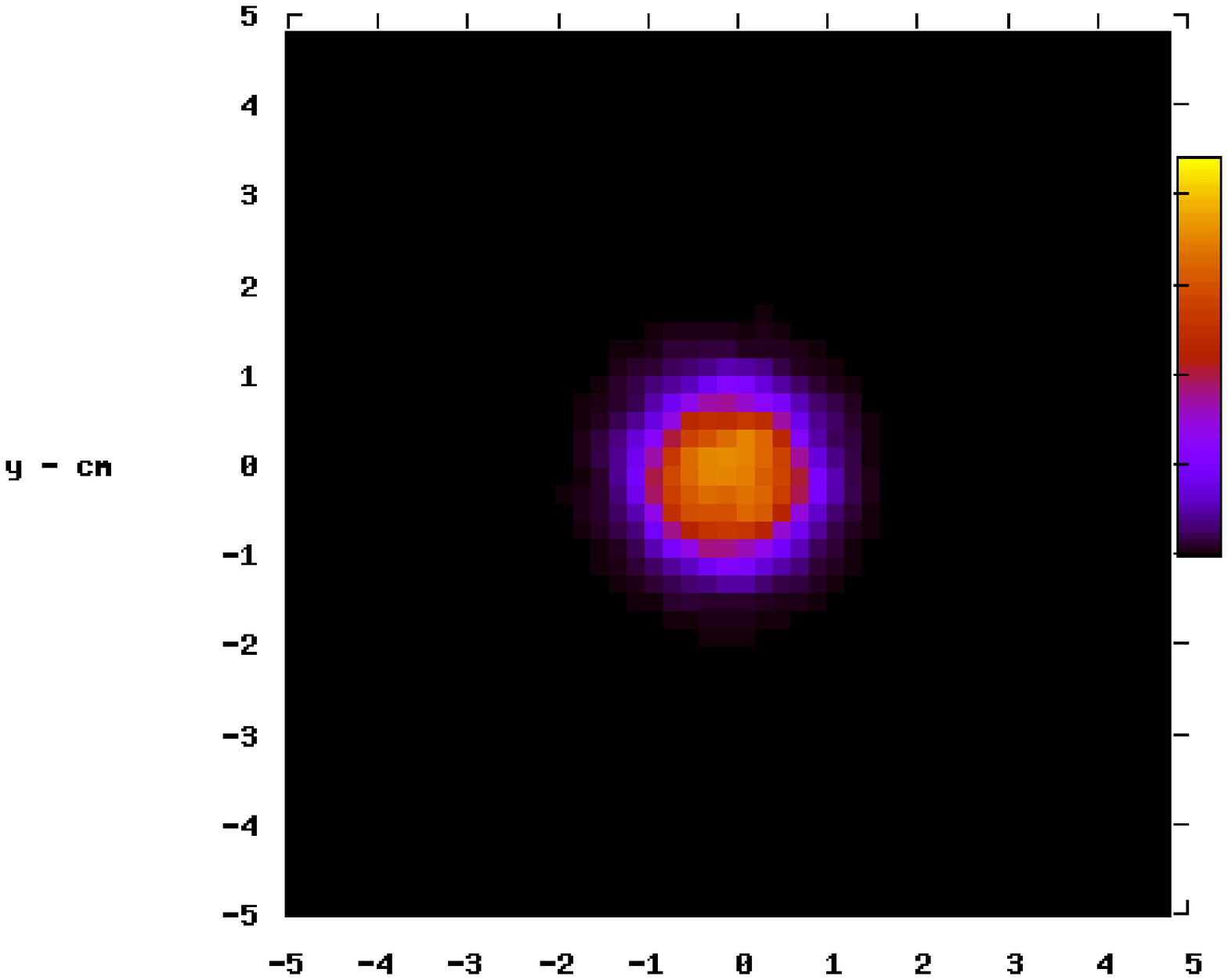}}
\end{center}
\vspace{-0.5cm}
\caption{On axis response function (PSF) of lenses made of flat mosaic crystals. 
{\em Left panel:} lens with 40 m focal length for an on-axis source in the 150--600 keV 
energy band. {\em Right panel:} lens with 6 m focal length 
for an on-axis source in the 70--300 keV energy band. See text. Reprinted from 
Refs.~\citenum{Frontera08a,Frontera08b}.}
\label{f:psf_onaxis}
\end{figure}

In Fig.~\ref{f:psf_onaxis} we show the calculated on--axis Point Spread Functions for two cases:
a ring--shaped lens of 40 m focal length and a spiral--shaped lens of 6 m focal length. 
In the first case the lens has a 150--600 keV energy passband and a crystal tile cross section of
10$\times$10~mm$^2$,  while in the second case it has a 70--300 keV passband and a 
crystal tile cross section of 15$\times$15~mm$^2$. 
In both cases it is supposed that the crystals, made of Cu(111), have a mosaic structure with 1 arcmin 
spread and  are perfectly oriented in the lens. 
The 6~m focal length lens has been proposed for a balloon experiment \cite{Frontera08b}. 

%
%

%
%
\begin{figure}
\begin{center}
\includegraphics[angle=0, width=0.6\textwidth]{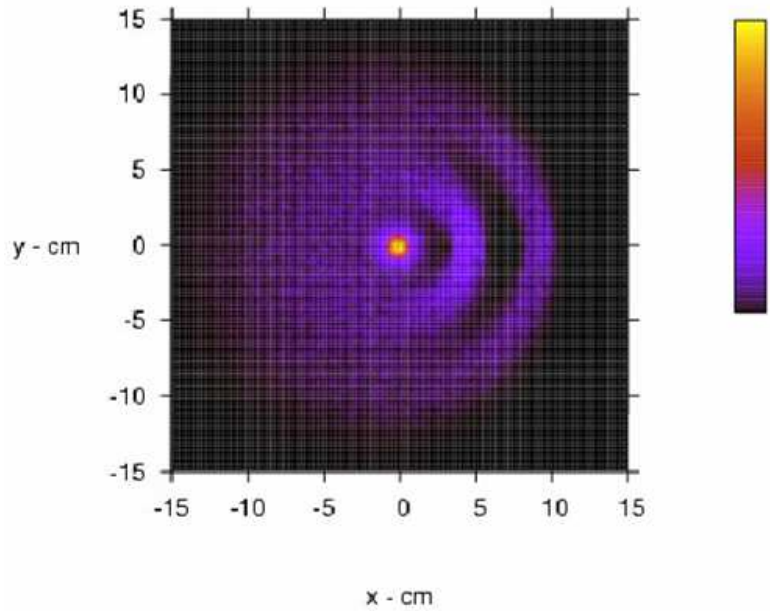}
\end{center}
\caption{Off-axis expected response function (PSF) of a Laue lens made of flat mosaic crystals
with 40 m focal length in the 150--600 keV energy band. 
Three point sources are simulated, at 0, 2 and 4 arcmin off--axis. See text.
Reprinted from Ref.~\citenum{Frontera08a}.}
\label{f:psf_offset}
\end{figure}

\begin{figure}
\begin{center}
\includegraphics[angle=0, width=0.8\textwidth]{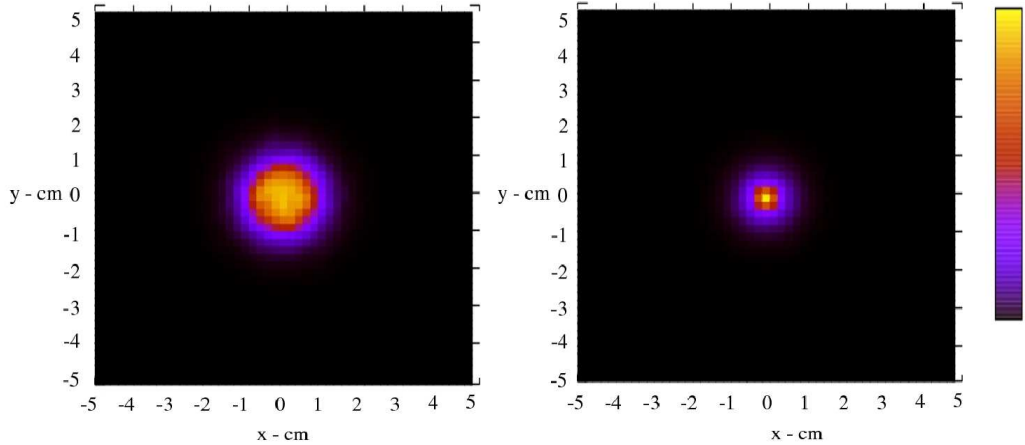}
\end{center}
\vspace{-0.5cm}
\caption{On axis response function (PSF) of a lens of 15 m focal length made of mosaic crystals 
with 30 arcsec spread, for an on-axis source. 
{\em Left panel:} Flat crystals with cross section of 15$\times$15~mm$^2$. 
{\em Right panel:} Curved crystals with a curvature radius of 30 m.}
\label{f:psf_curved}
\end{figure}

In the case of curved crystals, whose development is giving very satisfactory results (see
section~\ref{s:curved}), the expected PSF becomes very sharp, with a great advantage in terms of 
angular resolution and sensitivity. Assuming a lens of 15 m focal length made of crystals 
with a mosaic spread of 30 arcsec, the PSF obtained in the case of 15$\times$15~mm$^2$ flat mosaic 
crystals and that obtained in the case of curved crystals with a curvature radius of 30 m (2 times
the focal length) are shown in Fig.~\ref{f:psf_curved}. 
The difference between the two PSFs is outstanding. In the case of curved crystals  
we expect an angular resolution of 20 arcsec and a sensitivity higher than the corresponding
lens with flat crystals by  a factor of about 10.

%
%
In Fig.~\ref{f:psf_offset}, for flat crystal tiles of 15$\times$15~mm$^2$, we show the expected PSF 
of the 40 m focal length lens when three sources are in the Field of View (FOV), with one of the 
sources on--axis and the other two off--axis. As can be seen, in the case of off--axis sources the 
image has a ring shape centered in the on-axis source image, 
with inner radius that increases  with the offset angle, 
and a non uniform distribution of the reflected photons with azimuth. It is found that 
the integrated number of photons focused by the 
lens does not significantly vary from an on--axis source  to an off-axis source, but they are
spread over an increasing area. As a consequence, in principle
the FOV of the lens is determined by the detector radius, but, taking into account that
the focused photons from sources at increasing offset spread over an increasing area,
the lens sensitivity decreases with the source offset.
The azimuthal non uniformity of the PSF for off-axis sources can be usefully exploited, 
because it gives information on the azimuthal source direction.
The  angular resolution, in addition to the size of the flat crystal tiles, depends on the mosaic spread 
and on the misalignments of the lens crystal tiles. For the lens image shown in Fig.~\ref{f:psf_offset}, 
the angular resolution is of the order of, or even better than, 1 arcmin.  

\begin{figure}
\begin{center}
\includegraphics[angle=-90, width=0.4\textwidth]{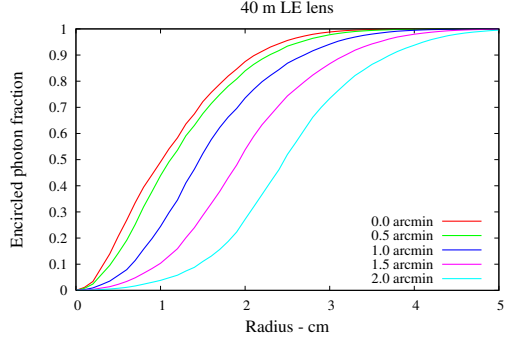}
\end{center}
\caption{Cumulative distribution of photons with the distance from the lens focus, for
different values of the mosaic spread, 40 m focal length lens, at about 100 keV.}
\label{f:encircled}
\end{figure}

Figure~\ref{f:encircled} shows the cumulative distribution of the on--axis photons with the
distance from the lens focus, for different values of the mosaic spread, in the case of
a 40 m focal length lens made of mosaic crystals of Cu(111) with 15$\times$15~mm$^2$ flat crystal tile cross section. 
As can be seen, for a low mosaic spread, the distribution is driven by the crystal size,
while for large spreads, it is mainly driven by this spread.  

%
%

\section{The development of Laue lenses}

Two key issues have to be faced in order build a Laue lens: 
\begin{itemize}
\item Development of technologies for the mass production  of suitable crystals 
(mosaic and/or curved crystals) in a reasonable time, consistent with that of 
preparation of a space mission;
\item Development of a technology for assembling, in an equally  reasonable time, 
thousands of crystal tiles in a lens with the 
proper orientation accuracy, that largely depends on the focal length.
\end{itemize}
Laue lens developments are being carried out in different institutions. We summarize here 
the major results obtained in our institutes.

\subsection{CLAIRE -- a narrow passband Laue lens experiment}

The objective of the R\&D project CLAIRE (French word that means 'clear')~\cite{Ballmoos05} 
was to demonstrate that
a prototype Laue lens can work under space conditions, measuring its performance by 
observing an astrophysical target. The CLAIRE telescope was flown twice (2000, 2001) on 
a stratospheric balloon by the French Space Agency CNES. CLAIRE's Laue lens was further 
tested on a 205 m long optical bench in 2003~\cite{alvarez04}. 
The project involved research groups from CESR (Centre d'E\'tude Spatiale des Rayonnements) Toulouse, 
University of Birmingham, Institute Laue-Langevin (ILL) Grenoble,
IEEC (Institut d'Estudis Espacials de Catalunya) Barcelona, and Argonne National Laboratory Chicago. 
CLAIRE's {\it narrow passband} lens consisted of 556 crystals 
(see Table 1) mounted on eight rings of a~45\,cm diameter Titanium frame. 
In each ring~i, the combination of the crystal plane spacing~$\mathrm{d}_{\mathrm{i}}$ and
the Bragg angle~$\theta_{\mathrm{Bi}}$ results in the concentration of
170\,keV photons onto a~common focal spot of 1.5\,cm diameter at
279\,cm behind the lens. 

CLAIRE's  crystals were produced by N. Abrosimov at the Institut 
f\"ur Kristallz\"uchtung (IKZ) in Berlin.  The Germanium-rich 
Ge$_{1-x}$Si$_x$ crystals (x $\approx$ 0.02) were grown by a modified 
Czochralski technique using Silicon feeding rods to replenish the loss of 
Si in the melt during the growth. The mosaicities of the Ge$_{1-x}$Si$_x$ crystals 
range between roughly 30 arcsec and 2 arcmin, leading to a field of view of 
about 1.5 arcmin and a diffracted energy bandwidth of about 3 keV at 170 keV. 
A correlative study between crystal structure, mosaicity and diffraction efficiency of 
the CLAIRE crystals is presented in Abrosimov et al (2005)\cite{Abrosimov05}. After 
cutting the crystal ingots at IKZ Berlin, most of the crystal tiles were 
characterized (mosaicity) at the Hard X-Ray Diffractometer of ILL Grenoble. 

At CESR Toulouse, the individual crystal tiles were 
mounted on flexible aluminium supports, which in turn are mounted on the lens frame. 
The reinforced 45 cm diameter titanium frame that holds up to 576 crystals on 8 rings  
was designed and manufactured at the Argonne National Laboratory, Argonne, USA.
The tuning of the lens consisted of tilting each crystal tile to the appropriate Bragg 
angle so that the diffracting energy was 170 keV for a source at infinity. Instead of 
directly calibrating the lens for a parallel beam of 170 keV photons, crystal tuning 
on the 20 m optical bench at CESR used a 150~kV X-ray generator situated on the lens 
optical axis at a distance of 14.16 m. At this distance, a crystal was correctly 
tuned (170 keV at infinity) if it diffracted 122.28  keV photons. A co-aligned mask 
(brass-lead sandwich) of the size of the entire lens, placed on the optical axis just 
in front of the lens, was used to select an individual crystal for tuning, while 
shielding already tuned crystals. 

\begin{table} [htb]  
\begin{center}       
\begin{tabular}{|c|c|c|c|c|c|c|}
\hline 
\rule{0pt}{2.5ex} 
ring & reflection & d [hkl] & size &number & radius & Bragg angle \\
 & [hkl] & [A] & [mm] & of crystals & [cm] & at 170 keV\\
\hline 
0 & 111 & 3.27 & 10 x 10 & 28 & 6.17 & 0.64 \\
1 & 220 & 2.00 &  10 x 10  & 52 & 10.08 & 1.04 \\
2 & 311 & 1.71 &  10 x 10  & 56 & 11.82 & 1.22 \\
3 & 400 & 1.41 &  10 x 10  & 72 & 14.26 & 1.48 \\
4 & 331 & 1.30 &  10 x 7  & 80 & 15.62 & 1.61 \\
5 & 422 & 1.15 & 10 x 10 &  96   & 17.47 & 1.81 \\
6 & 333 & 1.09 & 10 x 7 &  96  & 18.82 & 1.92 \\
7 & 440 & 1 & 10 x 10 &  104 & 20.17 & 2.09 \\
\hline 
\end{tabular}
\end{center}
\caption{{\it The crystalline plane, the crystal size, the number of crystals 
per ring as well as the ring radius and the Bragg angle at 170 keV are listed 
for each ring of the prototype crystal lens.}} 
\label{ta:par1lens}
\end{table}

The resulting geometric area of the CLAIRE lens was
$511\,\mathrm{cm}^2$, the FOV and the
passband were $90^{\prime\prime}$ and $\sim 3\,\mathrm{keV}$,
respectively. The photons were focused onto a~small $3\times3$~array of
high-purity Germanium detectors, housed in a~single cylindrical
aluminum cryostat.  Each of the single Ge bars was an n-type coaxial
detector with dimensions of $1.5\,\mathrm{cm}\times
1.5\,\mathrm{cm}\times 4\,\mathrm{cm}$.  Focusing onto such a~small
detector volume already results in very low background noise. The CLAIRE
stabilization and pointing system was developed by the balloon
division of the French space agency CNES. 

%
%
\begin{figure}
\begin{center}
\includegraphics[angle=0, width=0.7\textwidth]{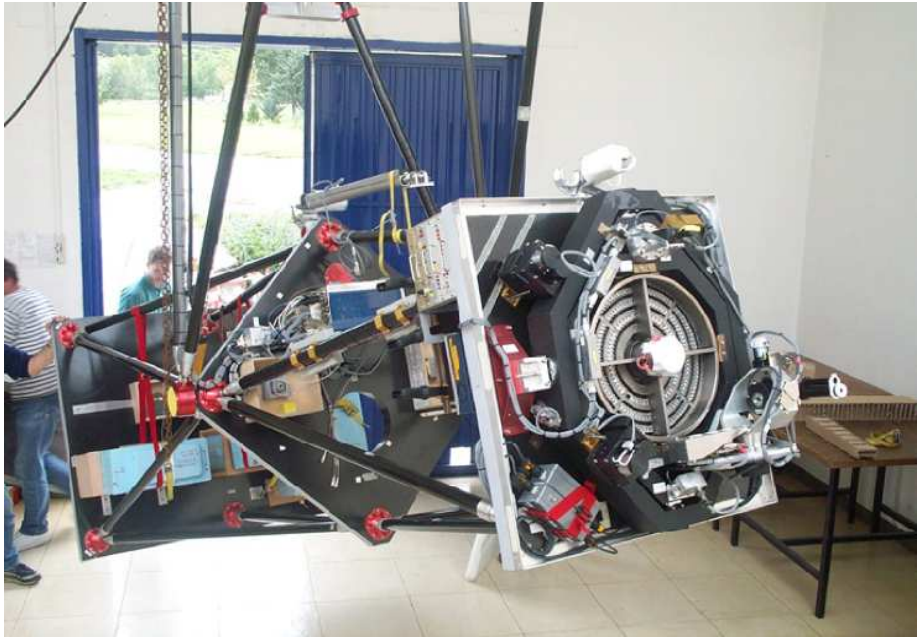}
\end{center}
\caption{The CLAIRE telescope  at the Gap--Tallard balloon base during the 2001 balloon campaign. On the first  
platform, the gamma-ray lens in its two-axes gimbal.}
\label{f:claire}
\end{figure}

\begin{figure}
\begin{center}
\includegraphics[angle=0, width=0.7\textwidth]{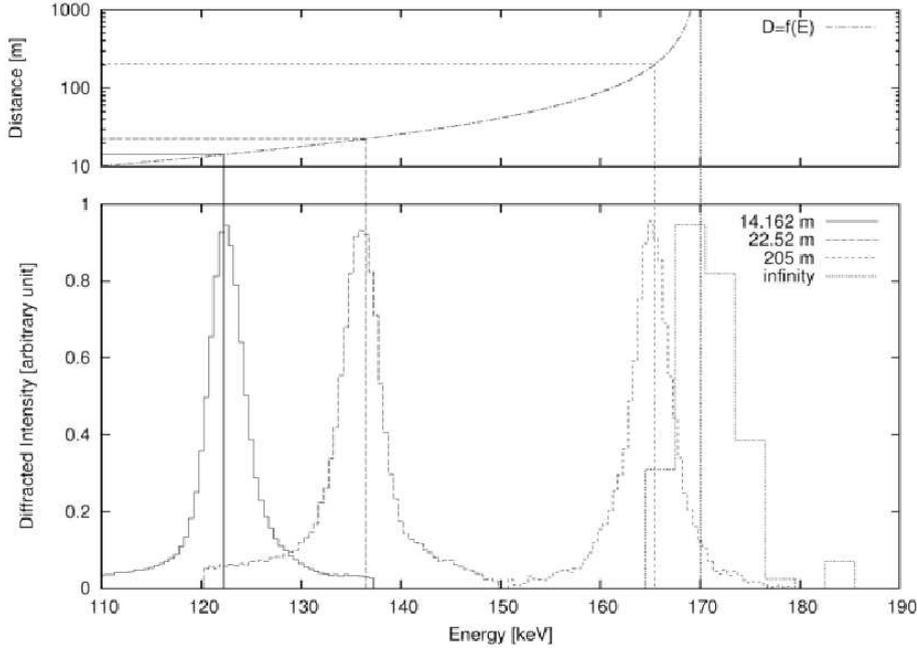}
\end{center}
\caption{Recorded spectra for various source distances (lower graph). 
The upper graph represents the distance of the source as a function of 
diffracted energy. The vertical lines show the theoretical corresponding energies. 
14.162 m corresponds to the tuning distance (E$_{th}$=122.29 keV). 
The measurement at 22.52 m (E$_{th}$=136.5 keV) was performed on the optical 
bench at CESR with a partially tuned lens. 205 m is the distance of the generator 
on the long distance test range (E$_{th}$=165.5 keV). The peak for an infinite 
distance is taken from the stratospheric flight of June 2001.}
\label{f:claire_diffpeak}
\end{figure}

{\bf CLAIRE's first light} : On June 14 2001, CLAIRE was launched by the balloon division of 
French Space Agency CNES from its base at Gap--Tallard in the French Alps (see Fig.~\ref{f:claire}). 
The astrophysical 
target was the Crab Nebula. (While nuclear lines are the perfect astrophysical targets for 
narrow passband Laue lenses, the balloon test flight ironically required  the observation 
of a~continuum spectrum.) With a mere 72 minutes of the flight having satisfactory pointing, 
CLAIRE nevertheless collected 33 photons from the Crab Nebula. This 3 $\sigma$ detection has 
been validated by ground tests conducted at distances of  14.16 m, 22.52 m (CESR optical bench) 
and 205 m (long distance test, Ordis, Catalunia).  
Figure~\ref{f:claire_diffpeak} shows the recorded spectra for these experiments. 
The energies of the centroids are in very good agreement with theory, slight departures 
from theoretical values (less than 0.5 keV) being the consequence of the incident 
spectrum shape and/or the detector calibration drifts. The measured peak efficiencies 
of the ground experiments are in fairly good agreement with the Crab observation: when 
their efficiencies are rescaled for a polychromatic source at infinity, a peak efficiency 
of 9$\pm$1\% is obtained. At first glance, this figure may seem rather modest, however 
when considering the constraints on a compact balloon instrument, the result is actually 
very positive. The short focal length leads to a lens with the outermost rings occupied by 
crystals with high reflection orders $n$ (see eq. 1): the outermost rings 
([333] and [440] crystals) are roughly 4 times less efficient than the innermost rings 
([111] and [220] crystals). Note that the largest numbers of crystals are situated in the 
outer rings where efficiencies are unfortunately lowest. A future space instrument will 
allow longer focal lengths and hence only low order crystalline planes with the highest 
efficiencies would be used.
Also, the quality of the CLAIRE crystals was quite heterogeneous: the efficiency of the 
individual crystals within a ring varied by factors of 2 to 10, depending on the ring! 
However, the CLAIRE lens contained crystals as efficient as the Darwin model predicts: 
the best crystals of the lens showed peak efficiencies well above 20\%.

CLAIRE's balloon flight provided the first observation of an astrophysical source 
with a gamma-ray lens. In combination with the long-distance test on the ground, these 
results validate the theoretical models and demonstrate the principle of Laue lens. 
Moreover, CLAIRE's stratospheric flight represents a first demonstration of the Laue 
lens technology in a space environment.

%
%

\subsection{Current crystal development}
 
The crystal development status is discussed in detail in 
Refs.~\citenum{Barriere09,Rousselle09}.
Mosaic crystals of Copper are  developed and produced at ILL
\cite{Courtois05}, gradient composition crystals of Si$_{1-x}$Ge$_x$ alloy with $x$ increasing 
along the crystal growth axis (curved crystal) are produced at IKZ 
\cite{Abrosimov05}. Crystal developments for Laue lenses are being undertaken in Italy:
mosaic crystals at the CNR-IMEM Institute, and curved crystals at the Physics Department, University 
of Ferrara (see Fig.~\ref{f:Si_Guidi}).

\subsection{Current lens assembly technology development}

Currently new lens assembly technologies are being developed  at the 
University of Ferrara, for building broad passband (70/100-600 keV) Laue lenses, and
at CESR for building high energy narrow passband (800--900 keV) Laue lenses.

The crystal tile positioning accuracy in the lens
is the most critical issue. It depends on the mosaic spread and the focal length. 
Longer focal lengths require better positioning accuracies, thus the development
is more challenging for lenses working at the highest energies.

\subsubsection{Assembly technology status for broad passband Laue lenses}

A technology for assembling crystal tiles for a moderately short ($\le$10--15 m) focal length lens 
is at an advanced stage of development at the University of Ferrara
\cite{Frontera07,Frontera08a,Ferrari09}.
It does not require any mechanism for a fine adjustment of the crystal orientation once the
crystal is positioned in the lens frame. Using this technology, a first lens prototype 
with 6 m focal length has already been developed and tested. It makes use of mosaic crystals 
of Cu(111). 

The technique adopted  is described in detail in Refs.~\citenum{Frontera07,Frontera08a}.
It makes use of a counter-mask provided with holes, two for each crystal tile. 
Each tile is positioned on the countermask by means of two cylindrical pins, 
rigidly glued to the crystal tile, that are inserted in the countermask holes. 
The pin direction and the axis of the average lattice plane of each crystal tile 
have to be exactly orthogonal. The hole axis direction constrains the energy
of the photons diffracted by the tile, while the relative position of
the two holes in the countermask establishes the azimuthal orientation of the axis 
of the crystal lattice plane. This axis has to cross the lens axis.

Depending on the direction of the hole 
axes in the countermask, the desired geometry of lens can be obtained. 
In the case of a lens for space astronomy, the hole axes have to be
directed toward the center of curvature of the lens. 
In the case of the developed prototype, the hole axes were set parallel to
the lens axis for the quick test of the lens with an X--ray tube,
that provides a divergent X--ray beam.

Once all the crystal tiles are placed on the counter-mask, a
frame is glued to the entire set of the crystals. Then the lens frame, along
with the crystals, is separated from the counter-mask and from the pins.
In the case of the first prototype, instead of using a chemical attack in order to 
separate the countermask from the lens as foreseen in the project,
a mechanical separation was attempted.

The first developed prototype is made of a 36 cm diameter ring of 20 mosaic crystal tiles. 
The mosaic spread  of the used crystals ranges from $\sim 2.5$ to $\sim 3.5$ arcmin.
The tile cross--section is 15$\times$15 mm$^2$ while its thickness is 2 mm. 
The lens frame is made of carbon fiber composite with a total thickness of 1 mm.

\begin{figure}
\begin{center}
\includegraphics[angle=0, width=0.4\textwidth]{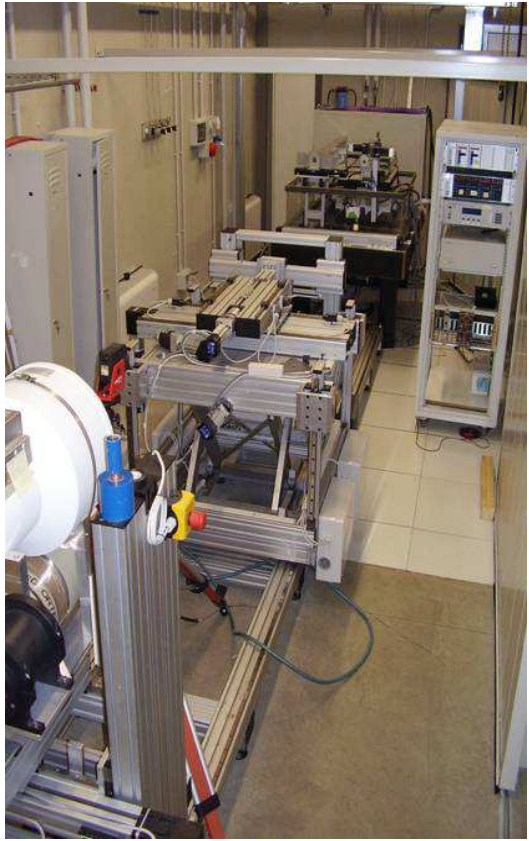}
\end{center}
\caption{A view of the current configuration of the apparatus for the lens 
assembling. The apparatus is located in the LARIX lab of the University
of Ferrara.}
\label{f:facility}
\end{figure}

The X--ray beam used, first to assemble the crystals and then to test the complete lens, 
is that available in  the LARIX (LArge Italian hard X--ray)
facility of the University of Ferrara. For a LARIX description see Ref.~\citenum{Loffredo05}. 
A view of the experimental apparatus in the current configuration 
is shown in Fig.~\ref{f:facility}. 
%
%

\begin{figure}
\begin{center}
\includegraphics[angle=0, width=0.4\textwidth]{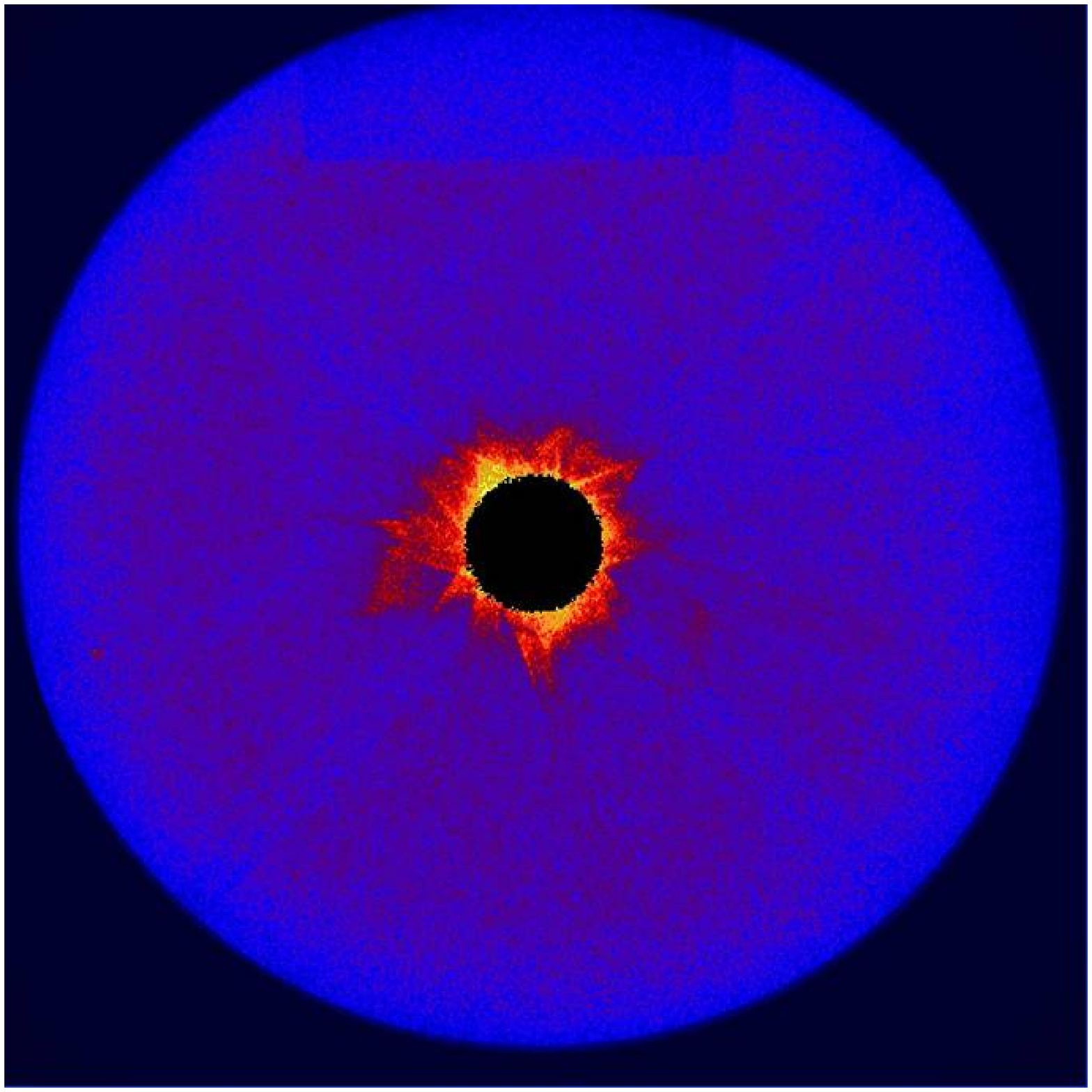}
\includegraphics[angle=0, width=0.4\textwidth]{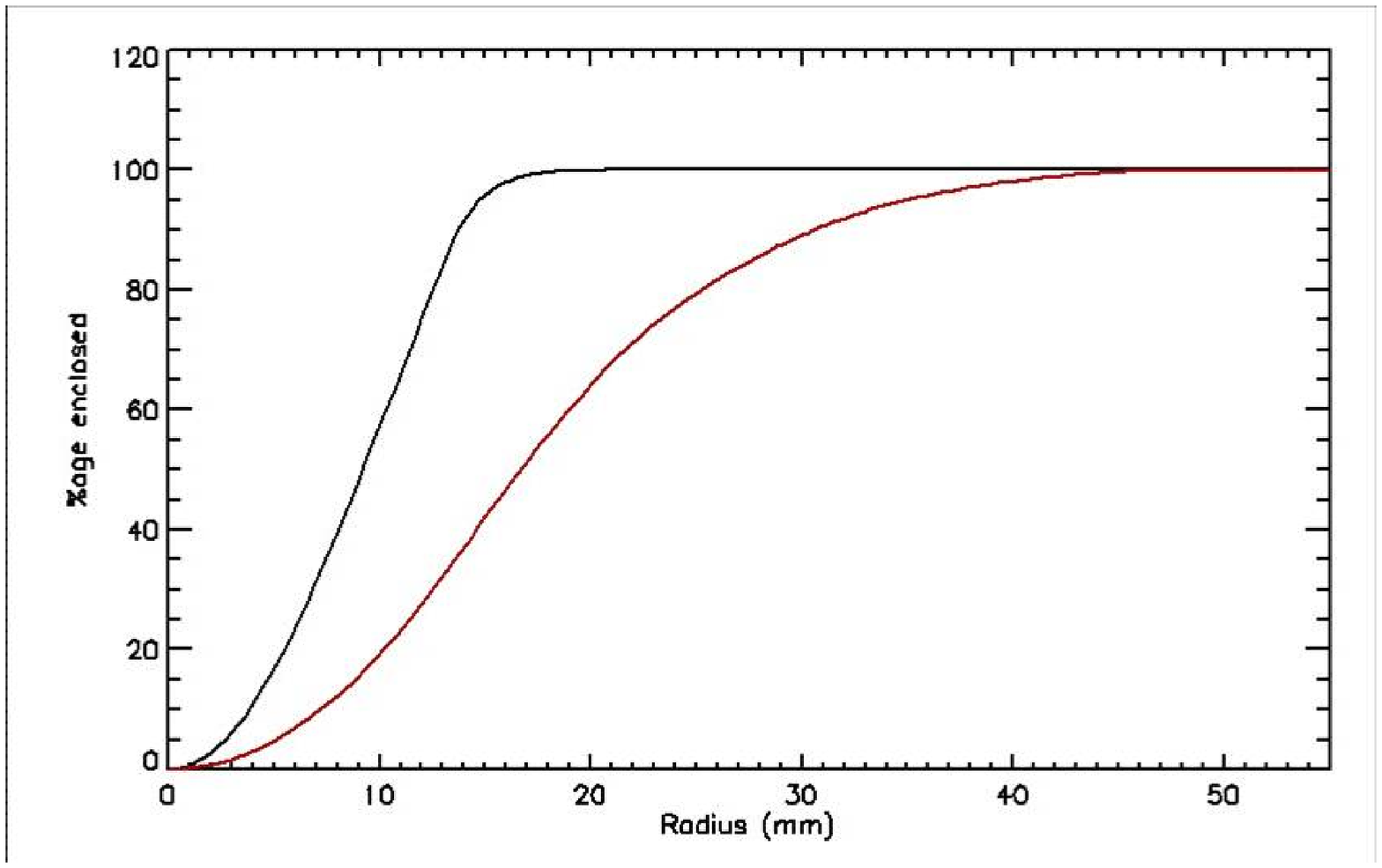}
\end{center}
\caption{{\em Left panel:} Difference between the PSF measured
and that obtained with a Monte Carlo code by assuming a perfect positioning 
of the crystal tiles in the lens. {\em Right panel:} cumulative distribution of the 
focused photons as a function of radial distance from the focal point. 
{\em Black line:} expected distribution in the case
 of a perfect positioning of the crystal tiles in the lens. {\em Red line:} measured distribution. 
Reprinted from Ref.~\citenum{Frontera08a}.}
\label{f:diff_image_spread}
\end{figure}

The prototype was thoroughly tested using the polychromatic X--ray beam described above, 
as reported in Ref.~\citenum{Frontera08a}.
Figure~\ref{f:diff_image_spread} shows the difference between the measured PSF 
and that obtained from a simulation, in which a perfect positioning of the crystals
in the lens was assumed.
As can be seen, only the center part of the measured image (i.e., the black region) 
is  subtracted by the simulated image. The corona still visible
in the difference image is the result of the cumulative error (mainly that due
to the mechanical separation of the lens from the countermask) during the lens assembly 
process.
%
%

The disagreement between the measured and the expected PSF is also apparent by comparing 
the cumulative distribution of the photons with the distance from the lens focus 
(see Fig.~\ref{f:diff_image_spread}, right panel). 
As can be seen, the PSF radius at which the expected fraction of focused photons 
reach the saturation (16 mm) corresponds to $\sim$60\% of the measured fraction.

\begin{figure}
\begin{center}
\includegraphics[angle=0, width=0.4\textwidth]{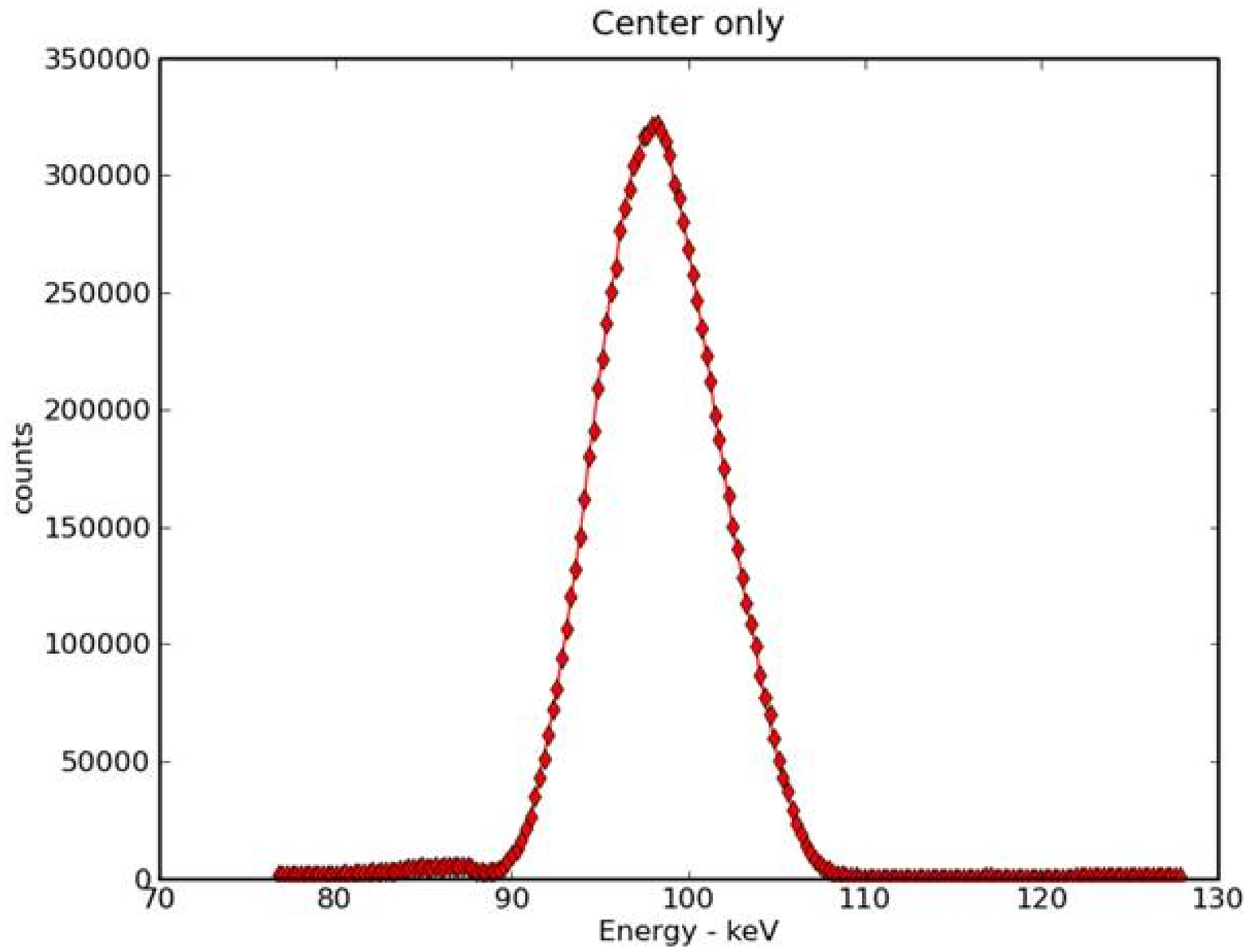}
\includegraphics[angle=0, width=0.4\textwidth]{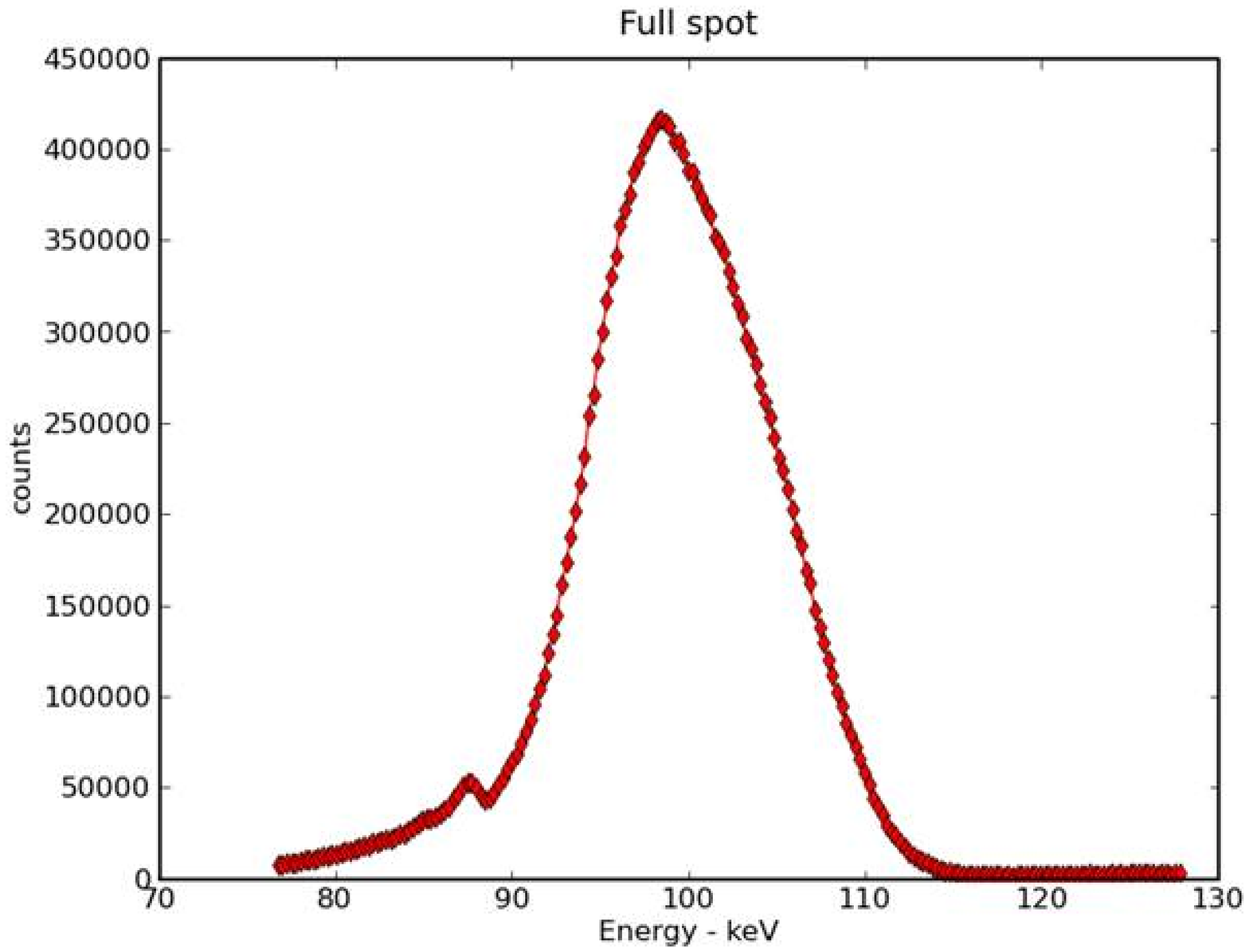}
\end{center}
\caption{{\em Left panel:} photon spectrum of the region (see 
left panel of Fig.~\ref{f:diff_image_spread}) in which all the reflected photons 
are expected to be found in the case of a perfect mounting of the 
crystal tiles in the lens. {\em Right panel:} spectrum of all reflected photons. Reprinted
from Ref.~\citenum{Frontera08a}. }
\label{f:spectrum}
\end{figure}

 The spectrum of the photons focused by the developed prototype is shown in Fig.~\ref{f:spectrum}, 
where we compare the measured spectrum of the central
region (i.e., photons in the black region of the left panel of Fig.~\ref{f:diff_image_spread})
with the spectrum of all reflected photons. As can also be seen 
from this figure, the centroid of the spectrum of the central region
achieves an intensity level 0.8 times that of the peak spectrum of all
reflected photons.
%
%

 A new prototype is being developed that takes into account the experience gained
 constructing the first one \cite{Ferrari09}.

\subsection{R\&D for a tunable narrow passband lens} 

\begin{figure}[htbp]\centering
\includegraphics[width=0.38\textwidth,clip]{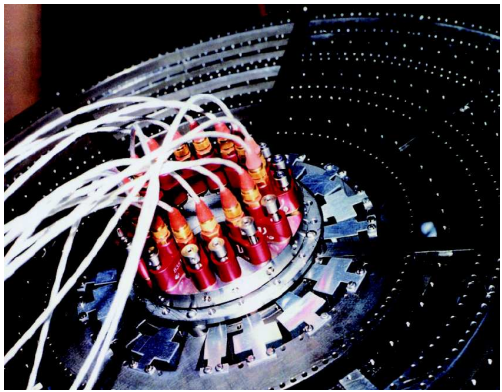}\hspace{8mm}
 \includegraphics[width=0.48\textwidth,clip]{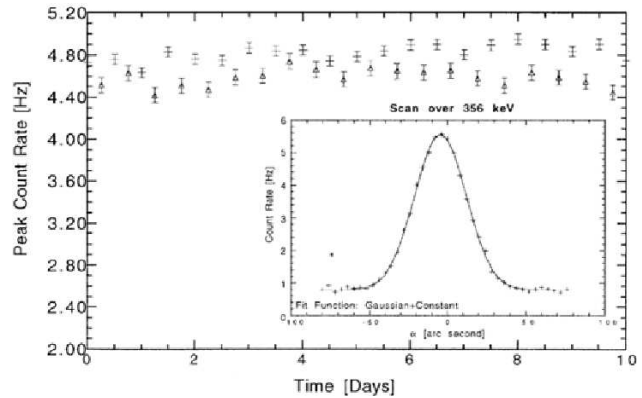}
\caption{a) Prototype tunable lens. b) The evolution in time of the peak 
count rate when alternatively focusing 303\,keV (circles) and 356\,keV 
(crosses) $\gamma$-rays demonstrates the stability and reproducibility of 
the lens tuning.}
\label{f:tunalens}
\end{figure}

Observing in \emph{only one}
narrow energy band might be considered too much of a handicap for a~space instrument. 
In the framework of an R{\&}D project for
the French Space Agency CNES, a tunable $\gamma$-ray lens prototype
(Fig.~\ref{f:tunalens} a) was developed and
demonstrated~\cite{Kohnle98}.  The capability to observe more than
one astrophysical line with a narrow passband Laue lens requires the tuning of 
two parameters: the Bragg angle~$\theta_{\mathrm{B}}$ and the focal length $f$. 
While the length $f$ will have to be controlled to within $\sim 1\,\mathrm{cm}$,
the precision of the crystal inclination has to be better than the
mosaic structure of the crystals. In the setup of Ref.~\citenum{Kohnle98}, each crystal 
was tuned by using piezo-driven
actuators to change the crystal inclination, and an eddy-current
sensor to determine the current position (Fig.~\ref{f:tunalens}a). The
resolution of the control-loop permitted an angular resolution of
0.1$-$0.4\,arcsec. The stability was found to be better than
0.8\,arcsec per day and the reproducibility of a~particular tuning
better than 5\,arcsec over a 10 day period (Fig.~\ref{f:tunalens}b).

%
%

\section{Prospects and conclusions}

A big effort has already been invested in the development 
of focusing Laue lenses for gamma--ray astronomy ($>$70/100 keV). Thanks to the most recent 
developments, Laue lenses with short focal length (10--15 m) are already feasible. The
major task now in progress is the development of the crystals needed to optimize the lens
effective area. A project "LAUE", supported by the Italian Space Agency, has just started 
in Italy (main contractor DTM, Modena) for the development of both suitable Laue lens crystals
and an advanced assembly technology for long focal length lenses (up to 100 m).

%
%
\begin{figure}
\begin{center}
\includegraphics[angle=90,width=0.4\textwidth]{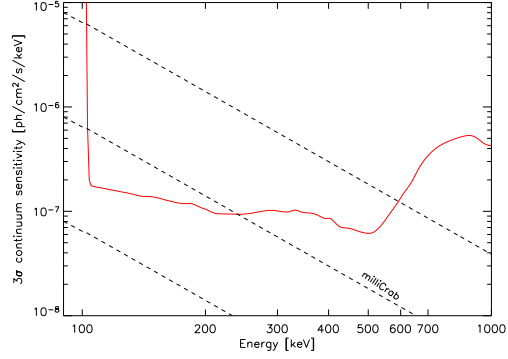}
\hspace{0.5cm}
\caption{3$\sigma$ continuum sensitivity ($\Delta E = E / 2$) of the lens 
discussed in Section 4.5.2 for 100 ks observation time. 
Reprinted from Ref.~\citenum{Barriere09c}.}  
\label{f:sens_Barriere09}
\end{center}
\end{figure}

Thus far, the major limit to the launch of a Laue lens gamma--ray telescope has been the need
of long focal lengths (20--100 m), that implies the use of two satellites in formation 
flying, one for the lens, and the other for the focal plane detector. The development of
extendable booms up to 20 m, the optimization of the lens effective area and the limitation
of the lens passband to lower energies, all make realistic the prospect of broad band
satellite missions that could join together multilayer mirrors and Laue lenses to extend the 
focusing band up to several hundreds of keV.

As an example, the 10$^5$ s continuum sensitivity of a Laue lens made of mosaic crystals,
that was investigated in
Ref. \citenum{Barriere09c} and mentioned above (see sect.~\ref{s:broad_band}), 
is shown in Fig.~\ref{f:sens_Barriere09}.  The use of curved crystals can further increase
the lens sensitivity.

%
%
\begin{figure}
\begin{center}
\includegraphics[angle=0,width=0.4\textwidth]{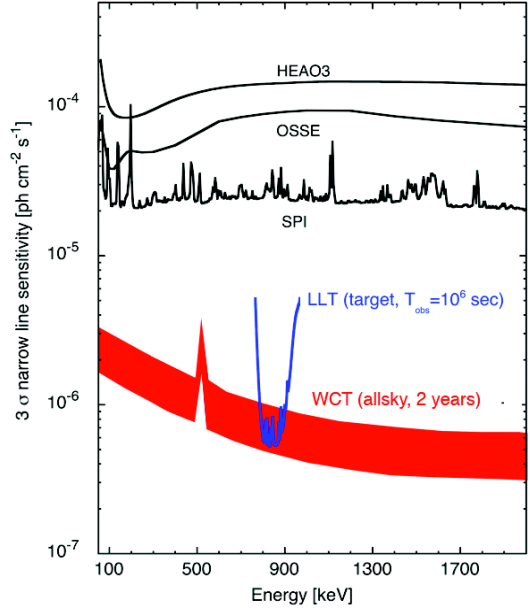}
\hspace{0.5cm}
\caption{Sensitivity estimation of DUAL's Laue Lens Telescope (LLT, blue curve), enabling study of SNe Ia 
out to ~50-80 Mpc, for dozens of potential targets each year. 
Also shown is the estimated survey sensitivity (two years) of the Wide Field Compton Telescope 
(WCT, red curve) which serves as a focal plane detector of the LLT.}
\label{f:dual_sens}
\end{center}
\end{figure}

In the framework of formation flying missions, 
the mission concept DUAL (Ref.~\citenum{PvB10}) is currently under study by a consortium of institutes 
from Europe, Japan and the USA. The DUAL mission is composed of a Wide-field Compton telescope 
(WCT) which carries out all-sky surveys, and a narrow band (800--900 keV) Laue-Lens Telescope (LLT) that 
simultaneously performs very deep observations of selected narrow-field targets. 
Combining a small Compton telescope with a Laue lens will permit the resolution of the apparently 
contradictory needs for a future gamma-ray mission: to map out large-scale distributions, 
monitor extreme accelerators, and measure the polarization of gamma-ray bursts with a 
small medium-sensitivity Compton Camera and simultaneously accomplish the stringent 
performance requirements for the specific science goal of  SN1a thanks to the Laue Lens. 
A  compact, wide field (2-3 $\pi$ steradian) Compton telescope with a modest geometric area 
(400-1600 cm$^{2}$) resulting in a effective area of 40-160 cm$^{2}$ can fulfill the needs 
for the "all-sky science" and simultaneously serve as a focal plane detector for the LLT. 
A Laue lens that, in $10^6$~s, can achieve sensitivities  of $10^{-6}$~photons cm$^{-2}$~s$^{-1}$ 
for a 3\% broadened line at 847 keV can be made from Au, Ag and Cu crystals, 
similar to those presently available. The model lens used for the sensitivity 
shown in Fig.~\ref{f:dual_sens} would have a focal length of the 
order of 68 m, a radius of  40-58 cm and a total mass of 40 kg. 

We expect that future broad band X--/gamma--ray missions for the deep study of  non 
thermal astrophysical processes above 100 keV, antimatter annihilation signatures, and nuclear lines 
from SN explosions will include Laue lenses.

\acknowledgments     
 
FF acknowledges the financial support by the Italian Space Agency ASI. 
PvB acknowledges continuing support from the French Space Agency CNES, and 
is particularly grateful to the balloon division of CNES which built the 
pointing system and operated the CLAIRE flights from the launch to the gondola recovery.
We also would like to thank the anonymous referees who have given valuable contributions
to improve our paper.


\bibliography{lens_biblio}   
\bibliographystyle{spiebib}   

\end{document}